\documentclass[a4]{article}

\usepackage[utf8]{inputenc}
\usepackage{subfiles}
\usepackage{xspace}
\usepackage{amstext,amsmath,amssymb}
\usepackage{graphicx}
\usepackage{array}
\usepackage{booktabs}
\usepackage[colorlinks=true,urlcolor=blue,linkcolor=blue,citecolor=blue,
  pdfborder={0 0 0}]{hyperref}
\usepackage[left=3cm,right=3cm,top=3cm,bottom=3cm]{geometry}
\usepackage[affil-it]{authblk}
\usepackage{cite}
\usepackage{multirow}

\newcommand{\particle}[1]{\ensuremath{\mathrm{#1}}\xspace}

\newcommand{\forcenewcommand}[1]{\providecommand{#1}{}\renewcommand{#1}}
\newcommand{\defparticle}[1]{
  \expandafter\forcenewcommand\csname #1\endcsname{\particle{#1}}
  \expandafter\forcenewcommand\csname #1bar\endcsname{\particle{\bar{#1}}}
}

\defparticle{e}
\defparticle{g}
\defparticle{q}
\defparticle{d}
\defparticle{u}
\defparticle{s}
\defparticle{c}
\defparticle{b}
\defparticle{p}
\defparticle{n}
\defparticle{N}
\defparticle{K}
\defparticle{D}
\defparticle{B}
\newcommand{\Tc}{\particle{\chi_{c1}(3872)}}
\newcommand{\Pc}[1][+]{\particle{P_c^{#1}}}
\newcommand{\Jpsi}{\particle{J/\psi}}
\newcommand{\etac}{\particle{\eta_c}}
\newcommand{\chic}[1][]{\particle{\chi_{\c#1}}}
\newcommand{\Lambdab}[1][0]{\particle{\Lambda_\b^{#1}}}
\newcommand{\Lambdac}[1][+]{\particle{\Lambda_\c^{#1}}}
\newcommand{\Sigmac}[1][+]{\particle{\Sigma_\c^{#1}}}

\newcommand{\DpiD}{\ensuremath{\D^0\,\pi^0\,\Dbar^0}\xspace}
\newcommand{\Jpsiomega}{\ensuremath{\Jpsi\,\omega}\xspace}
\newcommand{\Jpsirho}{\ensuremath{\Jpsi\,\rho^0}\xspace}
\newcommand{\DDstar}{\ensuremath{\D^0\,\Dbar^{*0}}\xspace}

\newcommand{\LambdacDstar}{\ensuremath{\Lambdac\,\Dbar^{*0}}\xspace}
\newcommand{\SigmacD}{\ensuremath{\Sigmac\,\Dbar^0}\xspace}
\newcommand{\SigmacDstar}{\ensuremath{\Sigmac\,\Dbar^{*0}}\xspace}
\newcommand{\SigmacstarD}{\ensuremath{\Sigmac[*]\,\Dbar^0}\xspace}
\newcommand{\SigmacstarDstar}{\ensuremath{\Sigmac[*]\,\Dbar^{*0}}\xspace}
\newcommand{\npi}{\ensuremath{\n\,\pi^+}\xspace}
\newcommand{\nrho}{\ensuremath{\n\,\rho^0}\xspace}
\newcommand{\ppi}{\ensuremath{\p\,\pi^0}\xspace}
\newcommand{\prho}{\ensuremath{\p\,\rho^0}\xspace}
\newcommand{\pomega}{\ensuremath{\p\,\omega}\xspace}
\newcommand{\petac}{\ensuremath{\p\,\etac}\xspace}
\newcommand{\pJpsi}{\ensuremath{\p\,\Jpsi}\xspace}
\newcommand{\pchic}{\ensuremath{\p\,\chic[0]}\xspace}
\newcommand{\pp}{\ensuremath{\p\,\p}\xspace}
\newcommand{\PcK}{\ensuremath{\Pc\,\K^-}\xspace}
\newcommand{\di}[1]{\ensuremath{#1^{+}\,#1^{-}}\xspace}

\newcommand{\plot}[1]{\includegraphics[width=0.49\textwidth]{#1}}
\newcommand{\ef}[2]{#1\scriptstyle\times10^{\scriptscriptstyle#2}}

\newcommand{\TeV}{\ensuremath{\mbox{TeV}}\xspace}
\newcommand{\GeV}{\ensuremath{\mbox{GeV}}\xspace}
\newcommand{\MeV}{\ensuremath{\mbox{MeV}}\xspace}
\newcommand{\nb}{\ensuremath{\mbox{nb}}\xspace}
\newcommand{\mb}{\ensuremath{\mbox{mb}}\xspace}
\newcommand{\ifb}{\ensuremath{\mbox{fb}^{-1}}\xspace}
\newcommand{\run}[1]{Run~{#1}\xspace}

\newcommand{\BR}{\ensuremath{\mathcal{B}}\xspace}
\newcommand{\pT}{\ensuremath{{p_\perp}}\xspace}
\newcommand{\CM}{\ensuremath{\mathrm{CM}}\xspace}
\newcommand{\AQM}{\ensuremath{\mathrm{AQM}}\xspace}
\newcommand{\eff}{{\mathrm{eff}}}
\newcommand{\tot}{{\mathrm{tot}}}
\newcommand{\res}{{\mathrm{res}}}
\newcommand{\el}{{\mathrm{el}}}
\newcommand{\inel}{{\mathrm{inel}}}
\newcommand{\bigo}[1]{\ensuremath{{\mathcal{O}({#1})}}}

\newcommand{\Pythia}{\textsc{Pythia}\xspace}

\renewcommand{\eqref}[1]{Equation~\ref{#1}}
\newcommand{\figref}[1]{Figure~\ref{#1}}
\newcommand{\tabref}[1]{Table~\ref{#1}}
\newcommand{\secref}[1]{Section~\ref{#1}}
\newcommand{\citeref}[1]{Ref.~\cite{#1}}

\newcommand{\etc}{\textit{etc.}\xspace}
\newcommand{\ie}{\textit{i.e.}\xspace}
\newcommand{\eg}{\textit{e.g.}\xspace}

\title{Forming Molecular States with Hadronic Rescattering}
\author{Philip Ilten}
\affil{
  Department of Physics,\\
  University of Cincinnati,\\
  Cincinnati, OH 45221, USA}
\author{Marius Utheim}
\affil{
  Theoretical Particle Physics,\\
  Department of Astronomy and Theoretical Physics,\\
  Lund University,\\
  S\"olvegatan 14A,\\
  SE-223 62 Lund, Sweden}
\date{}

\begin{document}

\sloppy

\begin{flushright}
  LU TP 21--31\\
  MCnet--21--13\\
  August 2021
\end{flushright}

\begin{minipage}[c]{\textwidth}
  \maketitle

  \vspace{\fill}

  \begin{abstract}
    \noindent
    A method for modelling the prompt production of molecular states
    using the hadronic rescattering framework of the general-purpose
    \Pythia event generator is introduced. Production cross sections of
    possible exotic hadronic molecules via hadronic rescattering at the
    LHC are calculated for the \Tc resonance, a possible
    tetraquark state, as well as three possible pentaquark states,
    $\Pc(4312)$, $\Pc(4440)$, and $\Pc(4457)$. For the \Pc states, the
    expected cross section from \Lambdab decays is compared to the
    hadronic-rescattering production. The \Tc cross section is compared
    to the fiducial \Tc cross-section measurement by LHCb and found to
    contribute at a level of \bigo{1\%}. Finally, the expected yields
    of \Pc production from hadronic rescattering during \run{3} of LHCb
    are estimated. The prompt background is found to be significantly
    larger than the prompt \Pc signal from hadronic rescattering.
  \end{abstract}

\end{minipage}

\thispagestyle{empty}

\clearpage 

\setcounter{page}{1}

\section{Introduction\label{sec:intro}}

While exotic bound quark states beyond the minimal $\q\qbar$ meson and
the $\q\q\q$ baryon structure have been proposed for some
time~\cite{Gell-Mann:1964ewy, Zweig:1964jf, Jaffe:1976ig,
  Strottman:1979qu, Lipkin:1987sk}, most experimentally observed
hadrons fit these minimal bound quark configurations. Prior to the
Large Hadron Collider (LHC), a number of observations for both exotic
tetraquark, $\q\q\qbar\qbar$, and pentaquark states,
$\q\q\q\q\qbar$, were claimed. However, many of these exotic states
could not be verified by later experiments~\cite{Hicks:2012zz},
excepting the $\particle{Z}(4430)$~\cite{Belle:2007hrb} and \Tc
resonances~\cite{Belle:2003nnu} which remain as possible tetraquark
candidates. Now with the LHC, more than $60$ new hadrons have been
observed with over $10$ new candidates for exotic tetraquark and
pentaquark states.

In 2015, the LHCb experiment discovered two resonances in the
$\Jpsi\,\p$ mass spectrum, which were identified as possible
pentaquark candidates, $\Pc(4380)$ and
$\Pc(4450)$~\cite{LHCb:2015yax}.\footnote{Inclusion of charge
  conjugate states and processes are implied throughout this work,
  unless explicitly noted in the text.} A subsequent 2019 LHCb
analysis with a larger data sample observed a possible additional
resonance, $\Pc(4312)$, and resolved the $\Pc(4450)$ pentaquark
structure as two resonances, $\Pc(4440)$ and
$\Pc(4457)$~\cite{LHCb:2019kea}. With this new observed mass
structure, a more in-depth amplitude study of the observed $\Pc(4380)$
resonance must be performed, leaving the existence of the $\Pc(4380)$
state ambiguous. The three viable \Pc candidates, $\Pc(4312)$,
$\Pc(4440)$, and $\Pc(4457)$, were not observed via prompt production
from the \pp collision point, but rather from the decay of \Lambdab
baryons

Similarly, the \Tc state (also known as $\particle{X}(3872)$) was
first observed through \B-meson decays by
Belle in 2003~\cite{Belle:2003nnu}, which was later confirmed by
BaBar~\cite{BaBar:2004oro}. More recently, LHCb measured the quantum
numbers of the \Tc to be $J^{PC} = 1^{++}$~\cite{LHCb:2013kgk,
  LHCb:2015jfc}. Lying within $0.2~\GeV$ of the \DDstar threshold, the
\Tc resonance is oftentimes interpreted as such a molecular
state~\cite{Tornqvist:2004qy}. The alternative interpretation of
the \Tc as a four-quark state is also a possibility, but the molecular
interpretation remains as the preferred model~\cite{Maiani:2004vq,
  BaBar:2004cah, Maiani:2017kyi, Bignamini:2009sk, Artoisenet:2009wk,
  Guo:2014ppa, Albaladejo:2017blx}. The $\Pc(4312)$, $\Pc(4440)$ and
$\Pc(4457)$ pentaquarks observed by LHCb may also be described by a
molecular state of \SigmacstarD or
\SigmacstarDstar~\cite{Chen:2015loa, Chen:2015moa, He:2015cea,
  Huang:2015uda, Roca:2016tdh, Lu:2016nnt, Shimizu:2016rrd,
  Shen:2016tzq, Yamaguchi:2017zmn}. Other models have also been
proposed for the \Pc states such as hadro-charmonium, a compact
charmonium state bound in light hadronic
matter~\cite{Dubynskiy:2008mq}.

The experimental observations above of exotic hadrons all consider
production from heavy hadron decays, $\B$-meson decays for the \Tc
state and \Lambdab for the \Pc states. Already, some predictions for
prompt production have been made using a coalescence type model where
free constituents of the molecular state may combine into a bound
molecular state if close in momentum space~\cite{Karliner:2004hk,
  Ling:2021sld}. These types of models have been successful in
modelling deuteron production at the LHC, including a full
implementation in \Pythia~\cite{Sjostrand:2014zea}, a general-purpose
event generator which allows for parameterised cross sections
differential in momentum space for multiple initial states. However,
these models do not consider the complete space-time picture of LHC
events, and require tuning of coalescence parameters to data, whether
cut-offs or overall normalisations.

Recent developments in \Pythia now allow hadronic resonances to be
formed from hadronic rescattering in a full space-time
picture~\cite{Sjostrand:2020gyg, Bierlich:2021poz}, where only the
partial widths of the hadronic resonance being formed are required to
fully specify the model. In this paper, this hadronic rescattering
framework has been modified to predict prompt exotic hadron production
for both the \Tc and \Pc states at the LHC. The details of the models
used to describe the exotic hadrons are introduced in
\secref{sec:model}, while results are given in \secref{sec:results}
and conclusions are given in \secref{sec:conclusion}.

\section{Models for exotic hadron production\label{sec:model}}

The hadronic rescattering framework of \Pythia can perform $2 \to 1$
scattering where the initial state hadrons combine to form a resonance
hadron. The cross section of this process depends on the mass and
total width of the resonance, as well as the partial width of the
given channel for the resonance. In the default \Pythia framework,
only pre-defined resonances and rescattering channels can be used for
rescattering. In this work, the framework has been expanded to allow
the addition of any arbitrary hadron resonance production from
rescattering. Specific configurations for \Tc, $\Pc(4312)$,
$\Pc(4440)$, and $\Pc(4457)$ are then defined, given model
assumptions, to determine the hadronic rescattering cross sections.

\subsection{Exotic hadron properties}

The \Tc mass is well measured to be $3871.69 \pm
0.17~\MeV$~\cite{ParticleDataGroup:2020ssz}. However, the \Tc widths,
both total and partial, are not as well known. In this work, the total
width of the \Tc is set to the world average, $1.19 \pm 0.21~\MeV$,
which is a combination of a dedicated inclusive LHCb line-shape
analysis~\cite{LHCb:2020xds} and a measurement by LHCb of \Tc
production from \B-decays~\cite{LHCb:2020fvo}. The partial widths of
the \Tc are set by normalising the central branching ratios, \BR,
reported by the PDG, and multiplying these by the total width. The
experimental uncertainty on these branching ratios is large, but the
\DDstar and \DpiD channels dominate. Because the latter is not a
two-body decay, the \Tc cannot be produced in rescattering through
this channel, but it still gives a significant contribution to the
total cross section. Likewise, the $\Jpsi \gamma$ and $\psi(2S)
\gamma$ are not used for resonance formation since photons are ignored
by the rescattering framework, but their contributions to the total
cross section are still included. The branching ratios and partial
widths used for the \Tc are given in \tabref{tab:widthsTc}.

\begin{table}
  \centering
  \caption{Experimentally measured branching ratios for the \Tc, as
    taken from the PDG~\cite{ParticleDataGroup:2020ssz}. The partial
    widths, in \MeV, for hadronic rescattering are given for each
    channel and are calculated as the product of the normalised
    branching ratio and the experimentally measured \Tc width of
    $1.19 \pm 0.21~\MeV$~\cite{ParticleDataGroup:2020ssz}.
    \label{tab:widthsTc}}
  \begin{tabular}{>{$}l<{$}|>{$}l<{$} >{$}l<{$}}
    \toprule
    & \textrm{PDG}~\BR & \textrm{width}~[\MeV] \\
    \midrule
    \mathrm{D}^0\,\bar{\mathrm{D}}^{*0}
    & \ef{(3.7\pm0.9)}{-1} & \ef{4.3}{-1} \\
    \mathrm{J}/\psi\,\omega
    & \ef{(4.3\pm2.1)}{-2} & \ef{5.0}{-2} \\
    \mathrm{J}/\psi\,\rho^0
    & \ef{(3.8\pm1.2)}{-2} & \ef{4.4}{-2} \\
    \mathrm{\chi_{c1}}\,\pi^0
    & \ef{(3.4\pm1.6)}{-2} & \ef{4.0}{-2} \\
    \mathrm{J}/\psi\,\gamma
    & \ef{(8.0\pm4.0)}{-3} & \ef{9.3}{-3} \\
    \mathrm{\psi}(2S)\,\gamma
    & \ef{(4.5\pm2.0)}{-2} & \ef{5.3}{-2} \\
    \mathrm{D}^0\,\pi^0\,\bar{\mathrm{D}}^0
    & \ef{(4.9^{~+~1.8}_{~-~2.0})}{-1} & \ef{5.7}{-1} \\
    \bottomrule
  \end{tabular}
\end{table}

The masses of the \Pc resonances are set to the central values of the
LHCb measurements. Experimental observations of the \Pc candidates are
limited to the \pJpsi decay channel, and so theory predictions based
on a molecular model from \citeref{Lin:2019qiv} are used instead to
define the total and partial widths. There, pentaquarks are treated as
\SigmacD/\SigmacDstar molecular states, with the $\Pc(4312)$ resonance
considered as a spin-1/2 \SigmacD state. The $\Pc(4440)$ and
$\Pc(4457)$ resonances are treated as \SigmacDstar states, and two
possible spin assignments are considered, either spin-1/2 or
spin-3/2. For all spin configurations, the predicted total widths for
the \Pc states are consistent with the observed widths, although these
widths have large experimental uncertainty. \citeref{Lin:2019qiv}
suggests that the $\Pc(4440)$ resonance is most likely spin-1/2 and
the $\Pc(4457)$ resonance is spin-3/2, but notes that the opposite
assignment cannot be excluded. Different form factors can also be
used, and so two different models are considered in this study. Model
1 uses the $(f_1, f_3)$ form factor set of \citeref{Lin:2019qiv},
while model 2 uses the $(f_2, f_3)$ set. Both models assume the
$\Pc(4440)$ resonance is spin-1/2 and the $\Pc(4457)$ resonance is
spin-3/2. The pentaquark partial widths used in this paper are
summarised in \tabref{tab:widthsPc}.

\begin{table}
  \centering
  \caption{Partial widths in \MeV for each pentaquark state of the two
    models considered from Ref.~\cite{Lin:2019qiv}. For both models
    the $\Pc(4440)$ is chosen to be spin-1/2 while the $\Pc(4457)$ is
    chosen to be spin-3/2.\label{tab:widthsPc}}
  \begin{tabular}{>{$}l<{$}|>{$}l<{$} >{$}l<{$} >{$}l<{$}
    |>{$}l<{$} >{$}l<{$} >{$}l<{$}}
    \toprule
    & \multicolumn{3}{c|}{model 1 width $[\MeV]$}
    & \multicolumn{3}{c}{model 2 width $[\MeV]$} \\
    & \mathrm{P_c^+}(4312) & \mathrm{P_c^+}(4440) & \mathrm{P_c^+}(4457)
    & \mathrm{P_c^+}(4312) & \mathrm{P_c^+}(4440) & \mathrm{P_c^+}(4457) \\
    \midrule
    \mathrm{\Lambda_c^+}\,\bar{\mathrm{D}}^0
    & \ef{6.0}{-2} & 5.6 & 1.5
    & \ef{3.0}{-1} & 2.7 & 1.2 \\
    \mathrm{\Lambda_c^+}\,\bar{\mathrm{D}}^{*0}
    & 3.8 & \ef{1.4}{1} & 6.1
    & \ef{1.1}{1} & \ef{1.2}{1} & 6.9 \\
    \mathrm{\Sigma_c^+}\,\bar{\mathrm{D}}^0
    & - & 3.4 & 1.0
    & - & 3.4 & \ef{9.0}{-1} \\
    \mathrm{\Sigma_c^{*+}}\,\bar{\mathrm{D}}^0
    & - & \ef{8.0}{-1} & 6.2
    & - & \ef{9.0}{-1} & 7.2 \\
    \mathrm{n}\,\pi^+
    & \ef{2.0}{-3} & \ef{1.0}{-3} & \ef{5.0}{-5}
    & \ef{8.5}{-1} & \ef{1.0}{-1} & \ef{3.0}{-1} \\
    \mathrm{n}\,\rho^+
    & \ef{2.0}{-5} & \ef{1.5}{-4} & \ef{1.0}{-5}
    & \ef{4.0}{-4} & \ef{2.0}{-1} & \ef{5.0}{-2} \\
    \mathrm{p}\,\pi^0
    & \ef{2.0}{-3} & \ef{1.0}{-3} & \ef{5.0}{-5}
    & \ef{8.5}{-1} & \ef{1.0}{-1} & \ef{3.0}{-1} \\
    \mathrm{p}\,\rho^0
    & \ef{2.0}{-5} & \ef{1.5}{-4} & \ef{1.0}{-5}
    & \ef{4.0}{-4} & \ef{2.0}{-1} & \ef{5.0}{-2} \\
    \mathrm{p}\,\omega
    & \ef{1.0}{-4} & \ef{1.0}{-4} & \ef{9.0}{-5}
    & \ef{3.0}{-3} & 1.5 & \ef{4.0}{-1} \\
    \mathrm{p}\,\eta_\mathrm{c}
    & \ef{1.0}{-2} & \ef{3.0}{-4} & \ef{6.0}{-5}
    & \ef{4.0}{-1} & \ef{7.0}{-2} & \ef{3.0}{-3} \\
    \mathrm{p}\,\mathrm{J}/\psi
    & \ef{1.0}{-3} & \ef{3.0}{-2} & \ef{1.0}{-2}
    & \ef{1.0}{-1} & \ef{6.0}{-1} & \ef{6.0}{-1} \\
    \mathrm{p}\,\mathrm{\chi_{c0}}
    & - & \ef{8.0}{-4} & \ef{3.0}{-5}
    & - & \ef{1.0}{-1} & \ef{3.0}{-3} \\
    \bottomrule
  \end{tabular}
\end{table}

\subsection{Exotic hadron production from rescattering}

In the hadronic rescattering framework of \Pythia, two hadrons will
interact if they pass each other in their centre-of-mass (\CM) frame
with an impact parameter $b < \sqrt{\sigma/\pi}$, where $\sigma$ is
the total cross section depending on the particle species and the \CM
energy. When two particles do interact, the specific process to
simulate is chosen with a probability proportional to the partial
cross section of that process. Since rescattering in \Pythia increases
the charged particle multiplicity, the recommendation of
\citeref{Sjostrand:2020gyg} is followed, where setting $\pT_{0}$
parameter of the multi-parton interaction (MPI) framework to
$2.345~\GeV$ compensates for this effect. This reduces the event
multiplicity before rescattering, with respect to the default \Pythia
MPI tune.

For processes involving charm hadrons, the total cross section in
\Pythia is calculated using the additive quark model
(\AQM)~\cite{Levin:1965mi,Lipkin:1973nt}. In this model, the total
cross section for two initial hadrons $A$ and $B$ is given by
\begin{equation}
  \sigma_{\AQM,\tot} = (40~\mb) \frac{n_{\eff,A}}{3}\frac{n_{\eff,B}}{3},
\end{equation}
where $n_\eff$ is the effective number of quarks in each hadron. In
\Pythia, this number is defined from the quark numbers of the hadron,
and by default is
\begin{equation}
  n_\eff = n_\u + n_\d + 0.6 n_\s + 0.2 n_\c + 0.07 n_\b.
\end{equation}
As an example, the total \pJpsi cross
section determined by the \AQM is $\sigma_{\AQM,\tot} = 5.33~\mb$. The
cross section for elastic scattering (in mb) is also determined with \AQM,
\begin{equation}
  \sigma_{\AQM,\el} = 0.039 \sigma_\AQM^{3/2}.
  \label{eq:aqmEl}
\end{equation}
The difference between the total \AQM cross section and the elastic
\AQM cross section gives an inelastic \AQM cross section of
\begin{equation}
\sigma_{\AQM,\inel} = \sigma_{\AQM,\tot} - \sigma_{\AQM,\el},
\end{equation}
which in default \Pythia corresponds to diffractive and
non-diffractive interactions.

In the model of \citeref{Lin:2019qiv}, pentaquarks can in principle
also form in \nrho, \prho, or \pomega interactions. \Pythia also uses
the \AQM model for the total cross section of these processes, but
pentaquark formation through these processes is so rare that the
contribution will be negligible. Finally, pentaquarks can be produced
in nucleon--pion interactions, \ie the \npi and \ppi channels. The
formation probability is also very small here, but now the
contribution may be non-negligible due to the abundance of these
particles in LHC collisions. For these processes, the total cross
sections at energies near the pentaquark masses are given by the
$\mathrm{HPR_1R_2}$ parameterisation~\cite{ParticleDataGroup:2018ovx}.

\begin{figure}
  \centering
  \plot{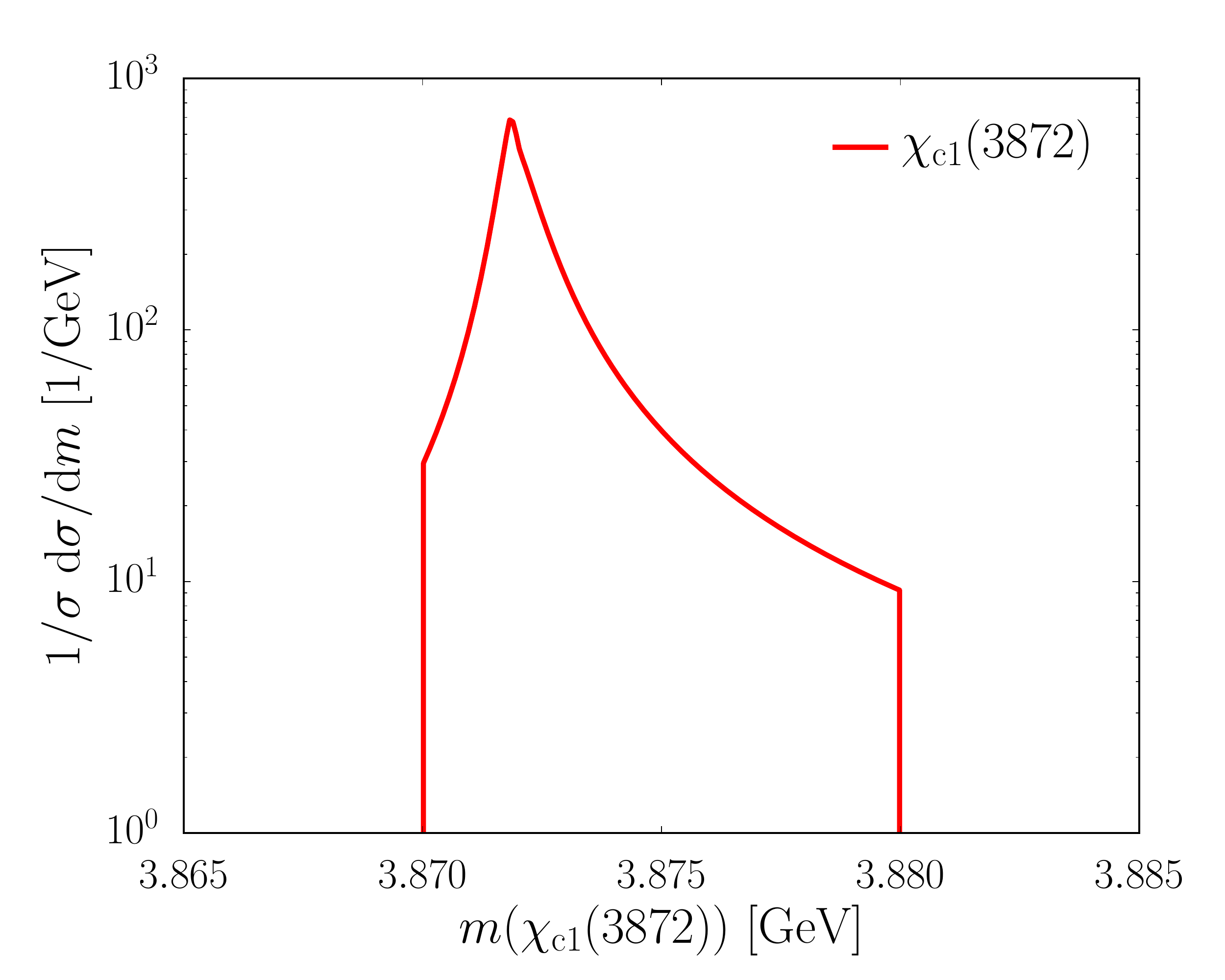}
  \plot{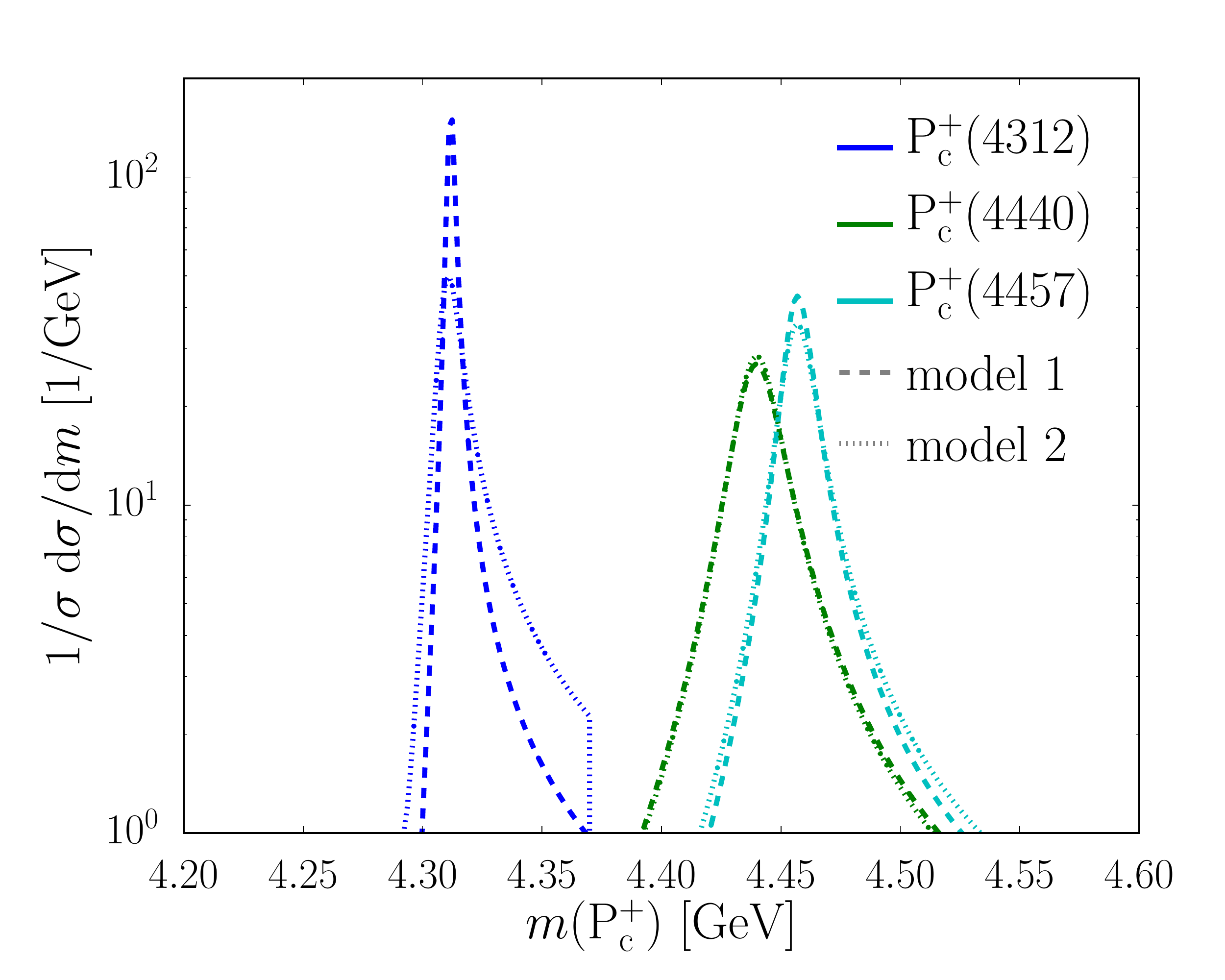}
  \caption{Mass distributions for the (left) \Tc tetraquark
    and (right) \Pc pentaquarks.\label{fig:mTcPc}}
\end{figure}

The partial cross section for a resonance formation process $AB \to R$
is given by a nonrelativistic
Breit--Wigner~\cite{ParticleDataGroup:2018ovx},
\begin{equation}
  \sigma_\res = \frac{\pi}{p_\CM^2} \frac{(2S_R + 1)}{(2S_A + 1)(2S_B
    + 1)} \frac{\Gamma_{R \to AB} \Gamma_R}{(m_R - E_\CM)^2 + \frac14
    \Gamma_R^2}, \label{eq:xsRes}
\end{equation}
where $p_\CM$ and $E_\CM$ are the momentum and energy of the incoming
particles in their \CM frame, $S$ is the spin of each particle, and
$m_R$ and $\Gamma_R$, $\Gamma_{R \to AB}$ are the mass, total width,
and partial width of the resonance, respectively. These widths are
mass dependent, as described in \citeref{Bass:1998ca}, giving mass
distributions as shown in \figref{fig:mTcPc}. Using these widths can
give mass distributions with longer tails than are physically
reasonable, and therefore explicit mass bounds are required. These
explicit cut-offs can give discontinuities in the mass distribution,
but this is not expected to significantly affect any relevant physical
observables.

It is important to keep in mind that resonance formation does not
change the mass spectrum when the decay products are the same as the
incoming particles. For instance, the process $\Lambdac \Dbar \to
\Pc(4440) \to \Lambdac \Dbar$ will not change the $\Lambdac \Dbar$
mass spectrum, as they must already be correlated in order to form the
resonance. However, in a system out of equilibrium, resonances can
change the relative composition of particles. If resonance production
receives a significant contribution from a particular channel, \eg
\ppi where the flux is large, but their decays are dominated by a
different channel, \eg $\Pc \to \Lambdac \Dbar$, then a peak structure
would be appear in that decay channel.

\begin{figure}
  \centering
  \plot{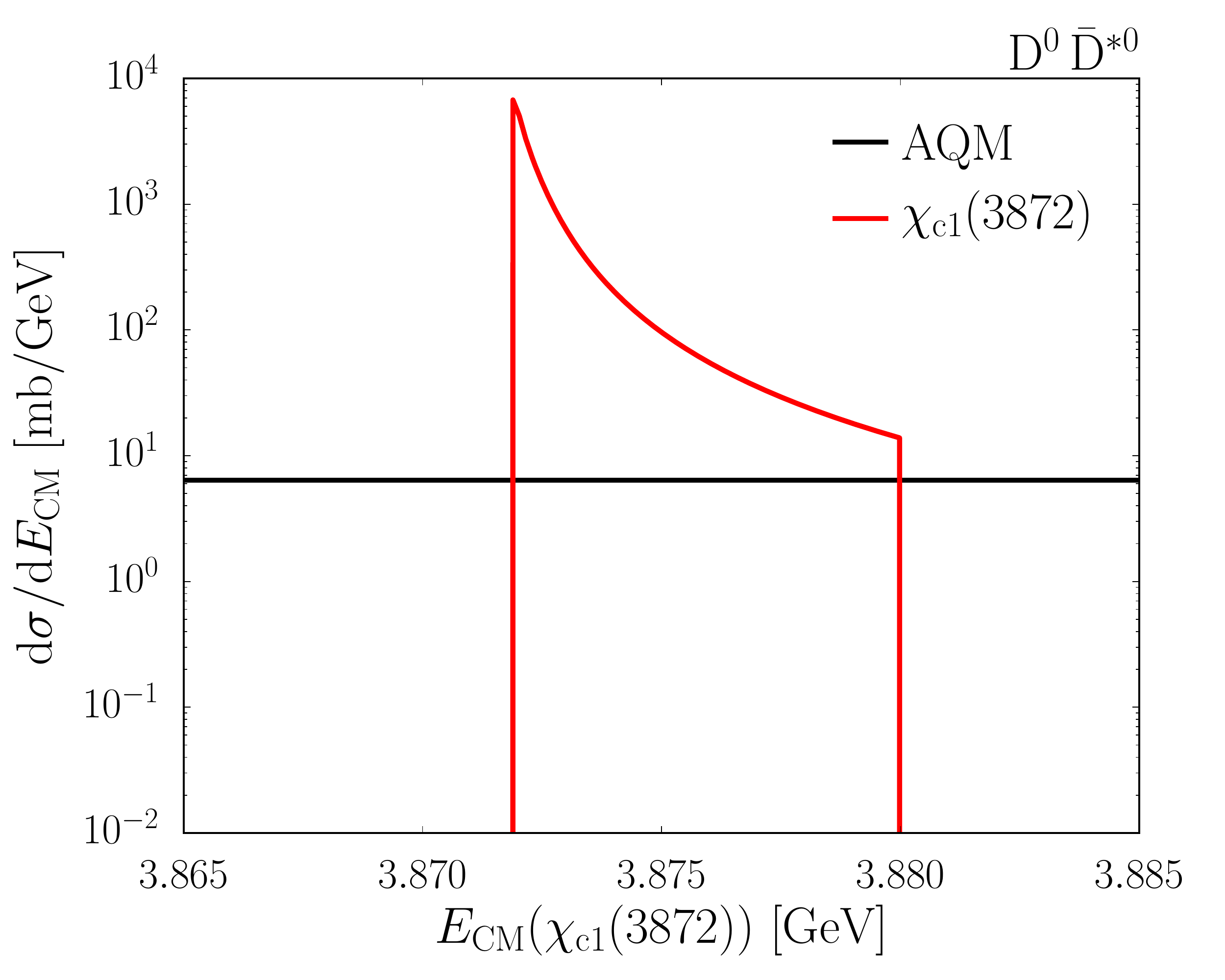} \plot{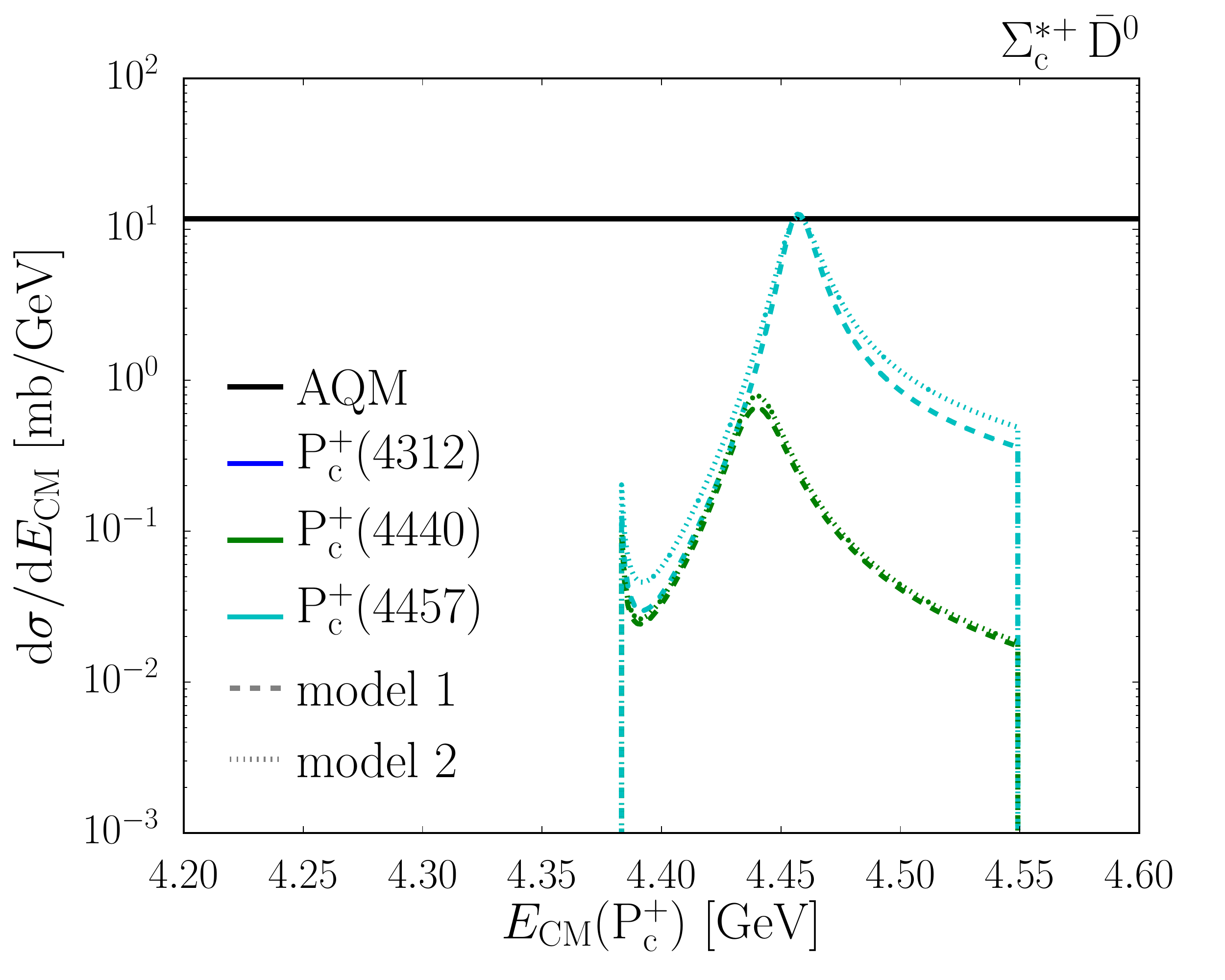}  \\
  \plot{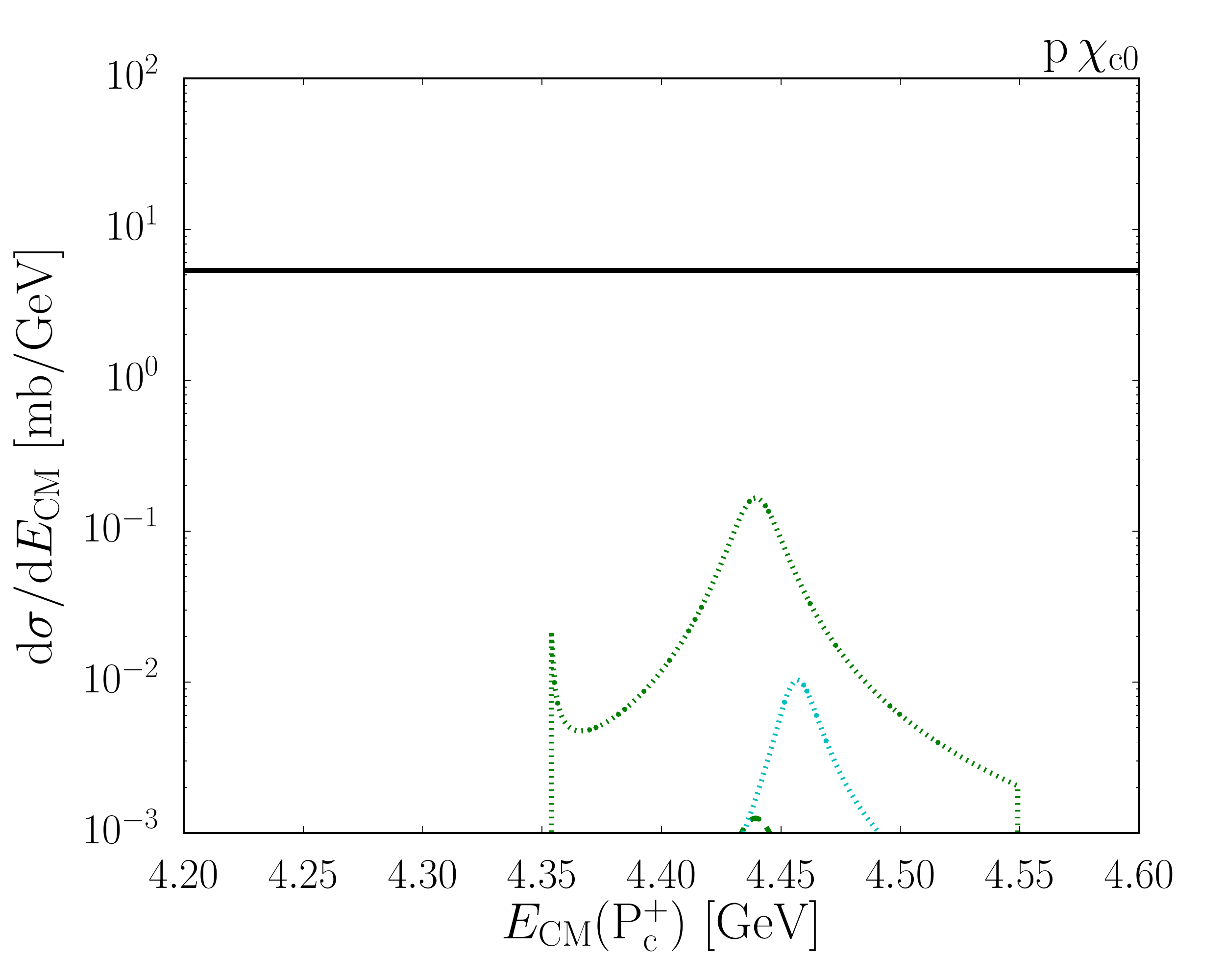}  \plot{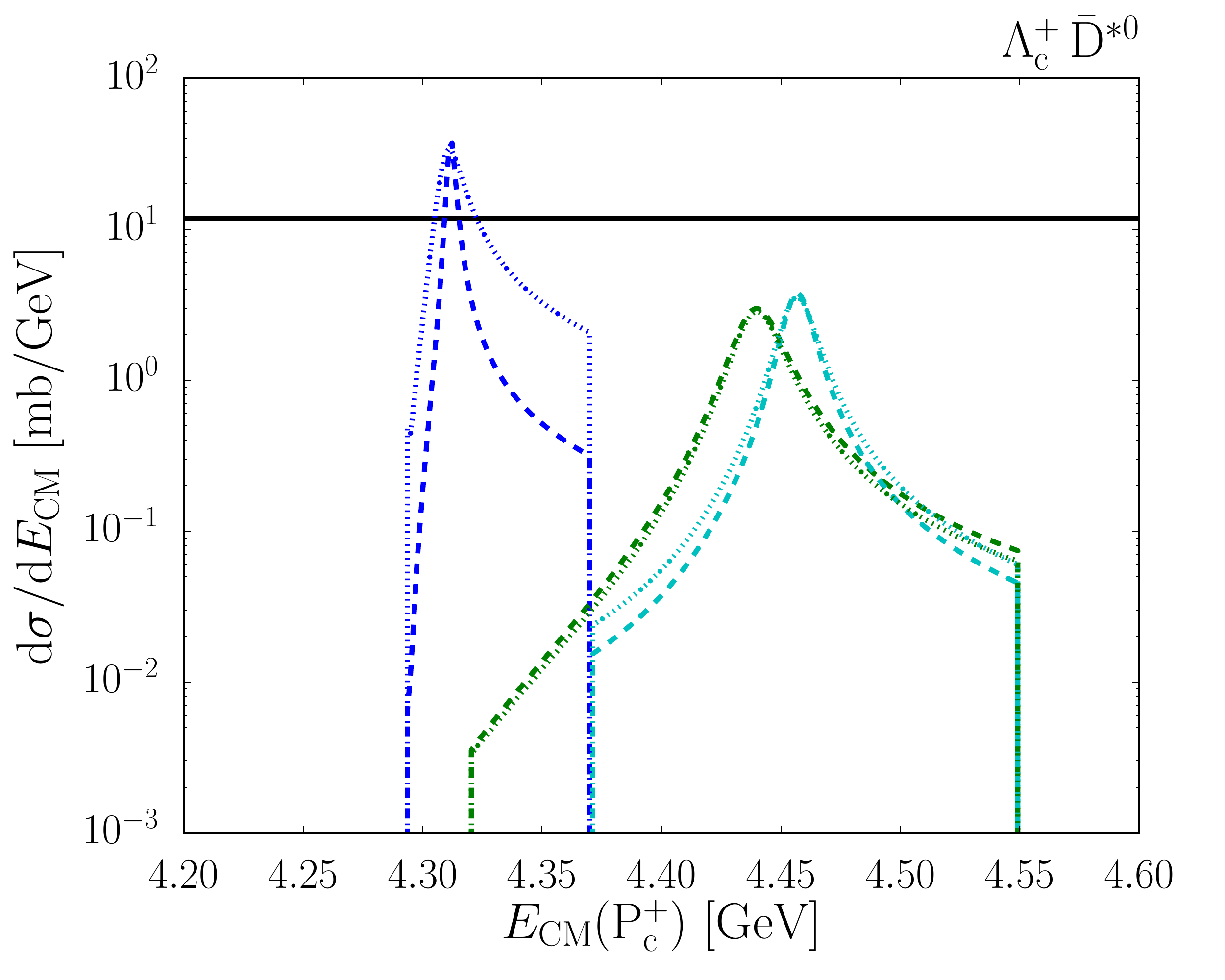} \\
  \plot{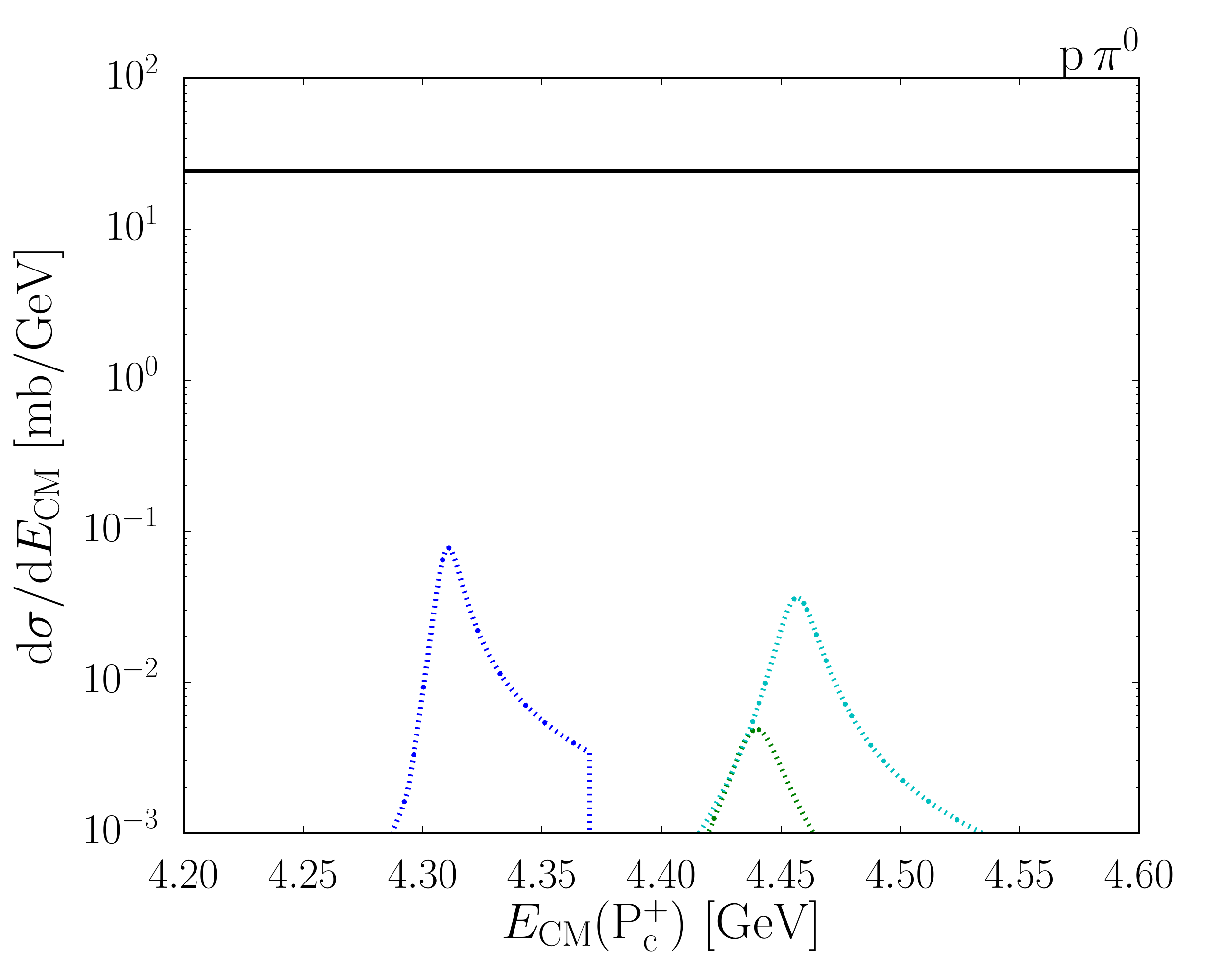}  \plot{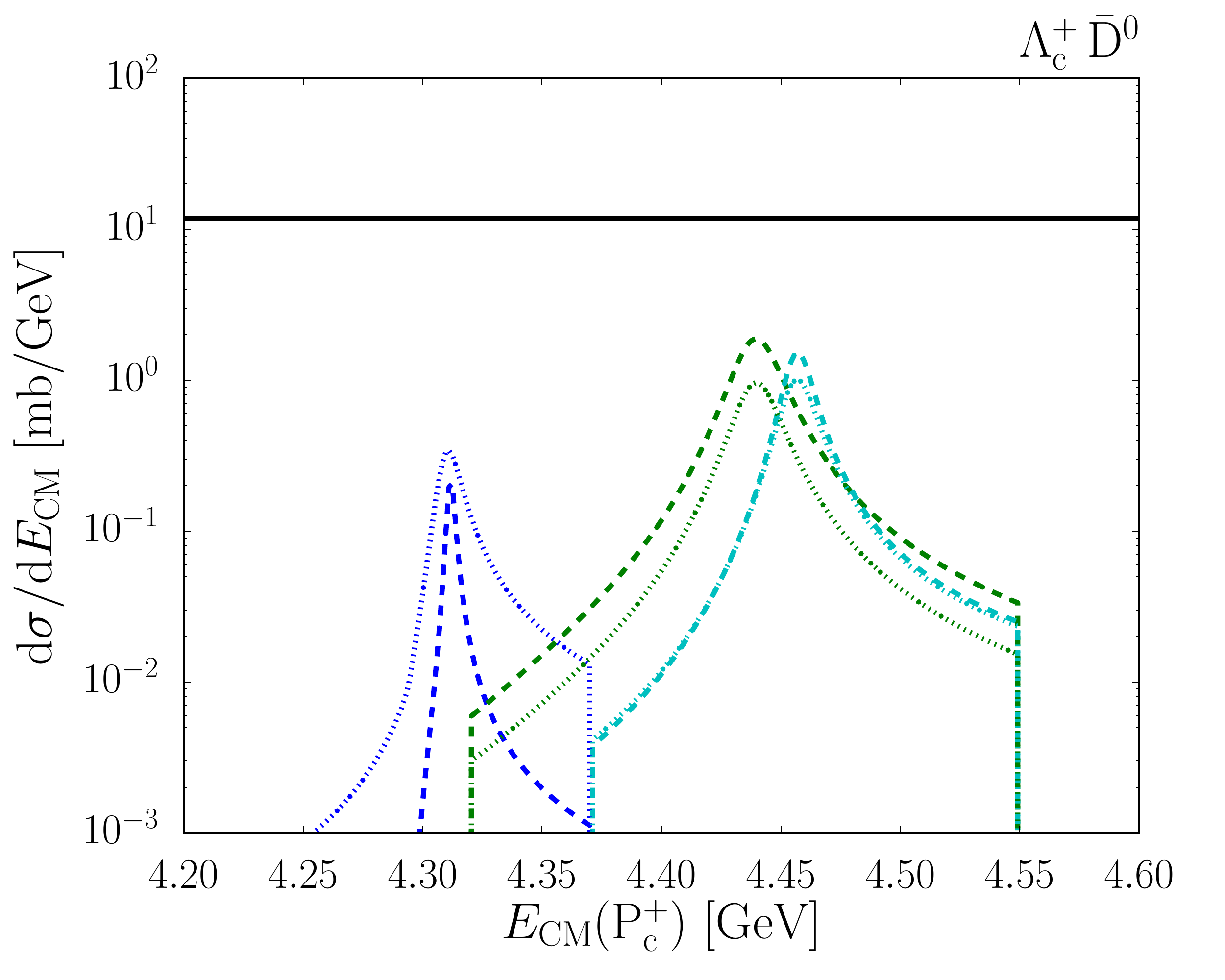}  \\
  \caption{Cross sections of the tetraquark and pentaquark resonance
    formation for relevant two-particle rescattering channels. The
    (solid black) total AQM cross section corresponds to the \Pythia
    default in the absence of exotic hadrons, and while drawn across
    the entire $E_\CM$ range is only available above mass threshold
    for each resonance. For the pentaquarks, cross sections are
    calculated using (dashed) model 1 and (dotted) model 2 given
    \tabref{tab:widthsPc}.\label{fig:xsECM}}
\end{figure}

When resonance formation is possible, the total cross section is
fixed, and $\sigma_{\AQM,\inel}$ is reduced by the resonance cross
section. If $\sigma_\res$ is greater than $\sigma_{\AQM,\inel}$, the
total cross section is increased to $\sigma_\res + \sigma_{\AQM,\el}$,
which as an example, occurs for the $\LambdacDstar \to \Pc(4312)$
process. The cross sections for tetraquark and pentaquark resonances,
as a function of the rescattering centre-of-mass energy, are shown in
\figref{fig:xsECM} for the primary rescattering channels. Some of
these cross sections grow very large near the kinematic threshold of
the channel, which is particularly visible for $\DDstar \to \Tc$.  The
technical reason for this is that the lower mass bound for the
particle lies below the threshold so the width does not vanish, hence
the factor $1/p_\CM^2$ in \eqref{eq:xsRes} dominates. Physically, this
can be motivated by the fact that slow-moving particles spend more
time near each other, and have a larger chance of interacting. For the
\Tc resonance, however, the cross section grows larger than what might
be considered reasonable considering the range of strong
interactions. In the rescattering framework, the range of interactions
is capped at a generous $5~\mathrm{fm}$, corresponding to a cross
section of roughly $785~\mb$, which limits the \Tc cross section. The
interpretation of such extremely large cross sections is not clear,
but a more detailed handling is outside the scope of this study.

\subsection{Pentaquark production from $\Lambdab$ decays}

While the focus of this study is exotic hadron production from hadronic rescattering, it is useful in the context of the \Pc states to compare these prompt production cross sections with the expected cross sections from \Lambdab decays. The branching ratios of the \Lambdab into \Pc states have not been experimentally measured, but can be fully determined given the partial widths of \tabref{tab:widthsPc} and the results of \citeref{LHCb:2019kea}. Here the contribution ratio is defined as
\begin{equation}
  \mathcal{R} =
  \frac{\BR(\Lambdab \to \PcK) \, \BR(\Pc \to \pJpsi)}
       {\BR(\Lambdab \to \pJpsi\,\K^-)},
\end{equation}
and has been measured for each \Pc resonance. The branching ratio in the
denominator is set as the experimentally measured value of
$\BR(\Lambdab \to \pJpsi\,\K^-) = (3.2^{+0.6}_{-0.5})\times 10^{-4}$~\cite{LHCb:2015qvk}.
By considering $\BR(\Pc \to \pJpsi)$ for each pentaquark state as set by
\tabref{tab:widthsPc}, the relevant \Lambdab branching ratios can be
determined by
\begin{equation}
  \BR(\Lambdab \to \PcK) = \BR(\Lambdab \to
  \pJpsi\,\K^-)\frac{\mathcal{R}}{\BR(\Pc \to \pJpsi)}.
\end{equation}
Values of $\mathcal{R}$ for each pentaquark state are also taken from
experiment~\cite{LHCb:2019kea} and are given in \tabref{tab:brsPc},
where the \Lambdab branching ratios are also provided. Note that in
model 1, the branching ratios for $\Pc \to \pJpsi$ are lower than for
model 2 by an order of magnitude or more, which gives a much smaller
$\Lambdab \to \PcK$ branching ratio for the former.

\begin{table}
  \centering
  \caption{Branching ratios for $\Lambdab \to \Pc\,\K^-$, determined
    using $\mathcal{R}$~\cite{LHCb:2019kea} and $\BR(\Lambdab \to
    \pJpsi\,\K^-)$~\cite{LHCb:2015qvk} from data, and $\BR(\Pc \to
    \pJpsi)$ calculated from \tabref{tab:widthsPc}. Both $\mathcal{R}$
    and $\BR(\Pc \to \pJpsi)$ are listed for each pentaquark state,
    where the individual uncertainties on $\mathcal{R}$ have been
    combined in quadrature.\label{tab:brsPc}}
    \begin{tabular}{>{$}l<{$}@{\hskip4pt} >{$}l<{$}|
      >{$}l<{$} >{$}l<{$} >{$}l<{$}}
      \toprule
      &
      & \mathrm{P_c^+}(4312)
      & \mathrm{P_c^+}(4440)
      & \mathrm{P_c^+}(4457) \\
      \midrule
      & \mathcal{R}
      & \ef{(3.0^{~+~3.5}_{~-~1.1})}{-3}
      & \ef{(1.1^{~+~0.4}_{~-~0.3})}{-2}
      & \ef{(5.3^{~+~2.2}_{~-~2.1})}{-3} \\
      \midrule
      \mbox{model 1} & \BR(\mathrm{\Lambda_b^0} \to \mathrm{P_c^+}\,\mathrm{K}^-)
      & \phantom{(}\ef{3.7}{-3}
      & \phantom{(}\ef{2.8}{-3}
      & \phantom{(}\ef{2.5}{-3} \\
      & \BR(\mathrm{P_c^+} \to \mathrm{p}\,\mathrm{J}/\psi)
      & \phantom{(}\ef{2.6}{-4}
      & \phantom{(}\ef{1.3}{-3}
      & \phantom{(}\ef{6.8}{-4} \\
      \midrule
      \mbox{model 2} & \BR(\mathrm{\Lambda_b^0} \to \mathrm{P_c^+}\,\mathrm{K}^-)
      & \phantom{(}\ef{1.3}{-4}
      & \phantom{(}\ef{1.3}{-4}
      & \phantom{(}\ef{5.1}{-5} \\
      & \BR(\mathrm{P_c^+} \to \mathrm{p}\,\mathrm{J}/\psi)
      & \phantom{(}\ef{7.6}{-3}
      & \phantom{(}\ef{2.7}{-2}
      & \phantom{(}\ef{3.4}{-2} \\
      \bottomrule
    \end{tabular}
\end{table}

\section{Results\label{sec:results}}

The cross sections of \figref{fig:xsECM} are not production cross
sections, but instead must be combined with the relevant flux of
initial state particles which can rescatter to produce molecular
states. High multiplicity environments are necessary to provide a
sufficiently large initial state flux, and so high energy hadronic
collisions, such as those produced at the LHC, are an ideal laboratory
to study possible molecular state formation from hadronic
rescattering. During \run{1}, the LHC collided proton-proton beams at
$\sqrt{s} = 7$ and $8~\TeV$, while during \run{2} $\sqrt{s} =
13~\TeV$, which corresponds to the majority of the LHC data
set. During \run{3}, the LHC is expected to run at a \CM energy of
$14~\TeV$, although $13~\TeV$ may also be used, depending upon the
performance of the collider. Consequently, a configuration with
$\sqrt{s} = 13~\TeV$ \pp collisions is conservatively chosen for this
study, since particle flux also increases as $\sqrt{s}$ is increased.

Using the default \Pythia parameter tune~\cite{Skands:2010ak} and a
modified version of \Pythia~8.306, the average visible final state
particle multiplicity for inelastic LHC events at $\sqrt{s} = 13~\TeV$
is expected to be \bigo{200}, with an inelastic cross-section of
$78~\mb$. This predicted inelastic cross-section is in good agreement
with LHC measurements~\cite{ATLAS:2016ygv, CMS:2018mlc}, including
forward measurements from LHCb~\cite{LHCb:2018ehw} and
TOTEM~\cite{TOTEM:2017asr}. The predicted particle density and energy
flow distributions also describe experimental LHC data
well~\cite{ATLAS:2017blj, CMS:2015zrm, LHCb:2011jir, LHCb:2012gpm},
across a number of experimental event categorisations intended to
separate elastic, diffractive, and inelastic scattering. Individual
particle species are also typically described well~\cite{LHCb:2012lfk,
  LHCb:2011ioc}, including open-charm meson
production~\cite{LHCb:2015swx}, although experimental measurements for
many rare mesons and baryons are not available for direct comparison.

The light pseudo-scalar mesons, $\pi^0$ and $\pi^\pm$, each have an
average multiplicity of \bigo{50}, while the average multiplicity for
light meson and baryons is at the level of \bigo{10} per species. This
includes the $\rho^0$, $\rho^\pm$, and $\omega$ vector mesons, and the
$\p/\pbar$ and $\n/\nbar$ light baryons. The $\D^0/\Dbar^0$ and
$\D^{*0}/\Dbar^{*0}$ mesons each contribute at an average multiplicity
of \bigo{10^{-1}}, while the \Lambdac[\pm] contributes at an average
multiplicity of \bigo{10^{-2}}. The relevant quarkonia states, \Jpsi
and \chic[0], have average multiplicities of \bigo{10^{-3}}, where
both colour octet and singlet contributions are included. For the
\Jpsi meson, feed-down production from both \chic-meson and \B-hadron
decays is also included. The \etac contribution is significantly less
at \bigo{10^{-4}}, but this is a known underestimation by \Pythia,
since production is included only through hadronization and not
through direct nonrelativistic QCD (NRQCD) calculations.

The production of both \Sigmac[\pm] and \Sigmac[*\pm] baryons is
relatively rare, with average multiplicities of \bigo{10^{-4}} and
\bigo{10^{-3}}, respectively. The production of
\Lambdab/$\bar{\Lambdab}$, relevant for displaced production of
pentaquark states, is also rare with an average multiplicity of
\bigo{10^{-3}}. Given these average multiplicities, the \DDstar
rescattering channel is expected to dominate \Tc production, while the
rescattering channels with \Lambdac baryons are expected to dominate
\Pc production. If the rescattering probability for the \Lambdac
channels, which depends on the kinematics of the scattering hadrons,
is similar to the $\Lambdab \to \PcK$ branching ratio, then \Pc
production via hadronic rescattering and \Lambdab decays is expected
to have roughly similar rates.

\subsection{Differential cross-sections}

\begin{figure}
  \centering
  \plot{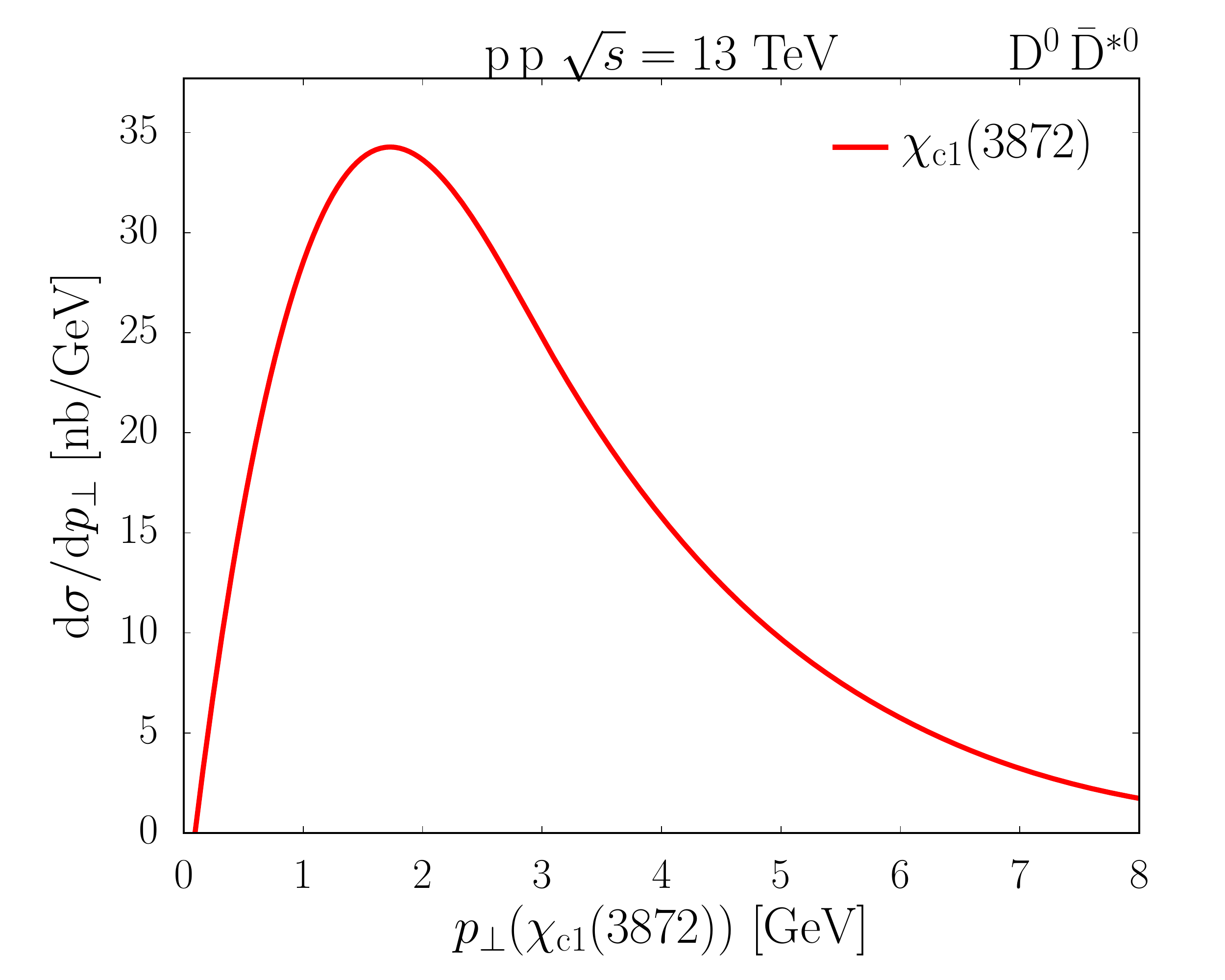} \plot{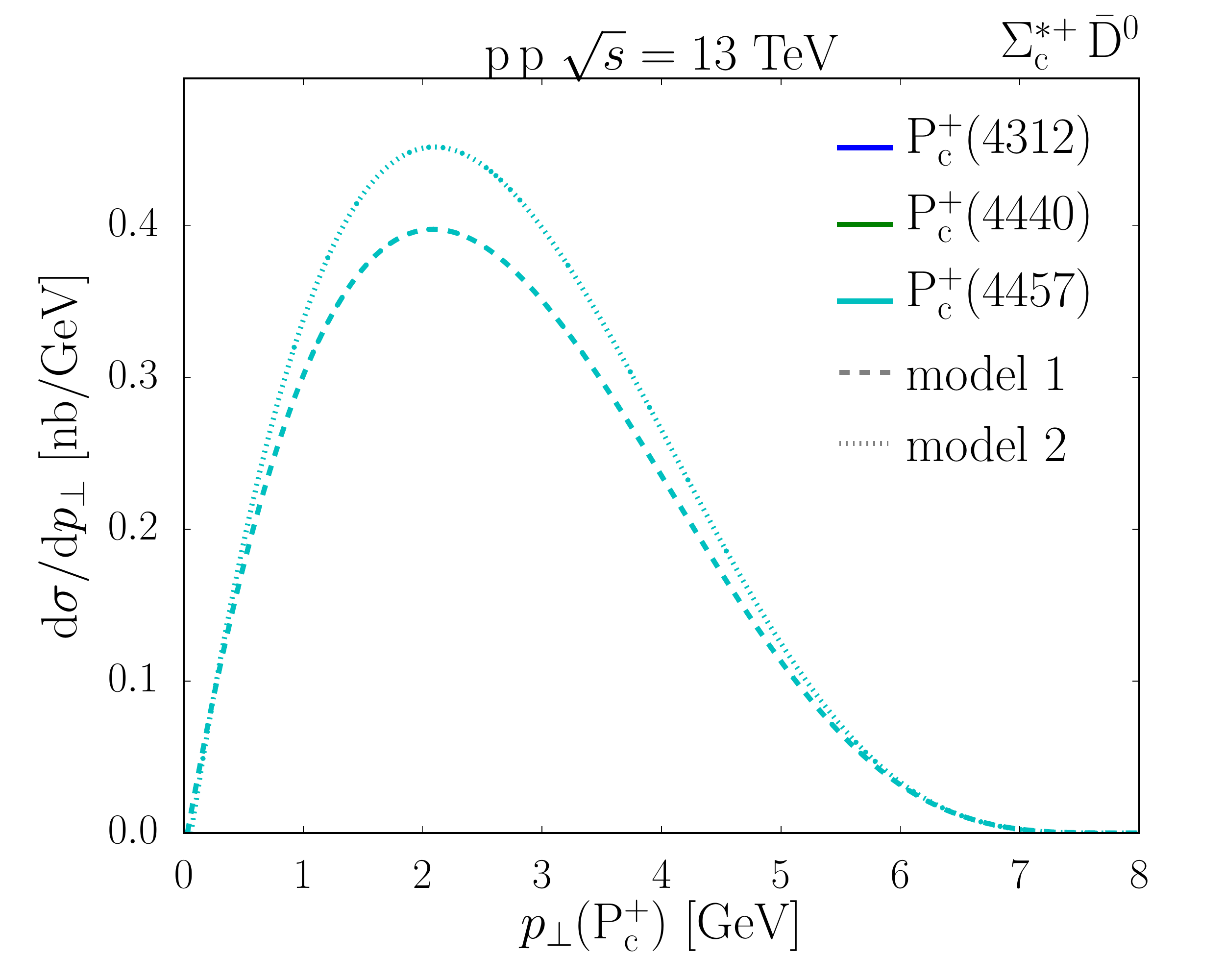}  \\
  \plot{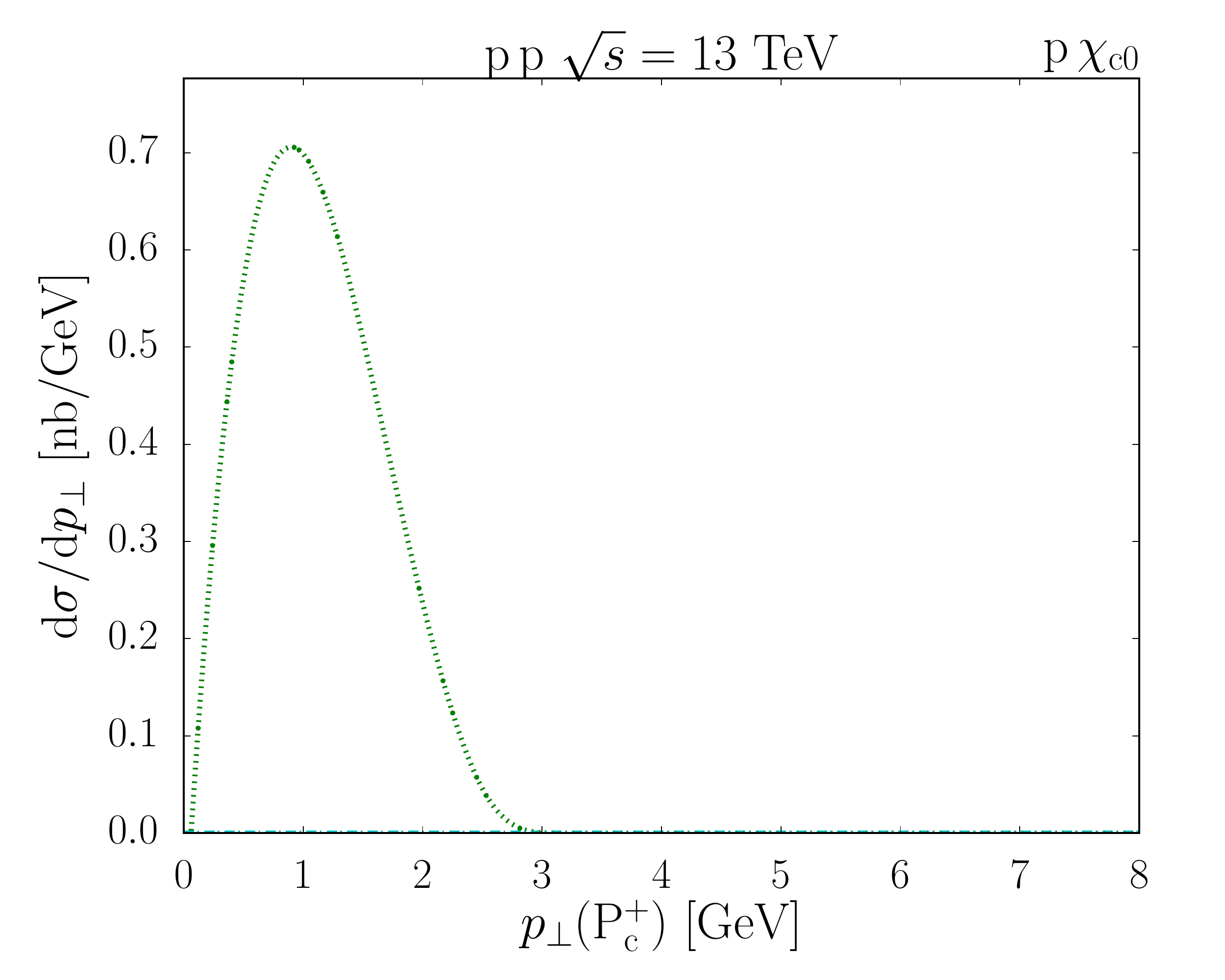}  \plot{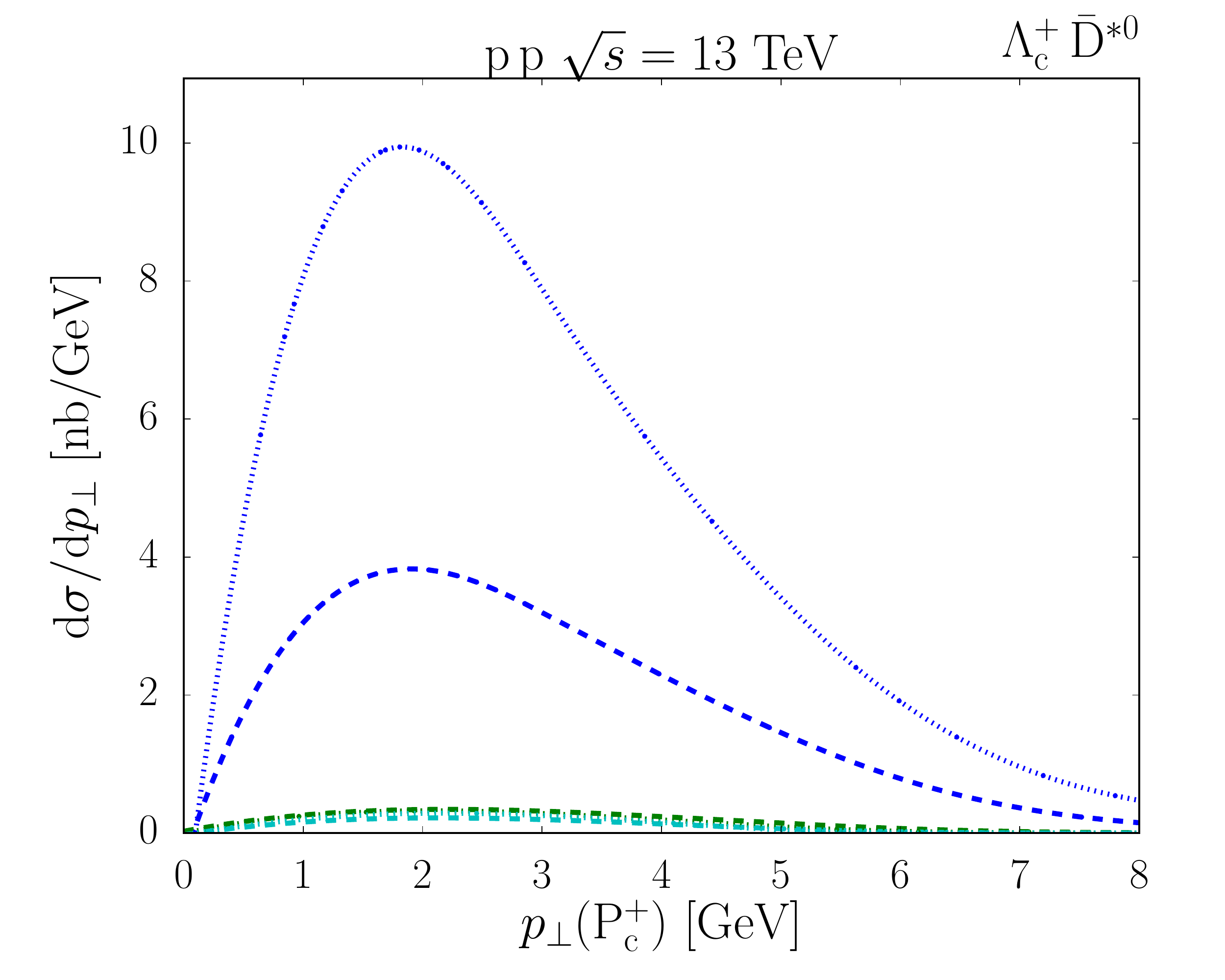} \\
  \caption{Tetraquark and pentaquark resonance hadronic-rescattering
    cross sections, differential in resonance \pT, produced in \pp
    collisions at $\sqrt{s} = 13~\TeV$.\label{fig:xsPt}}
\end{figure}

\begin{figure}
  \centering
  \centering
  \plot{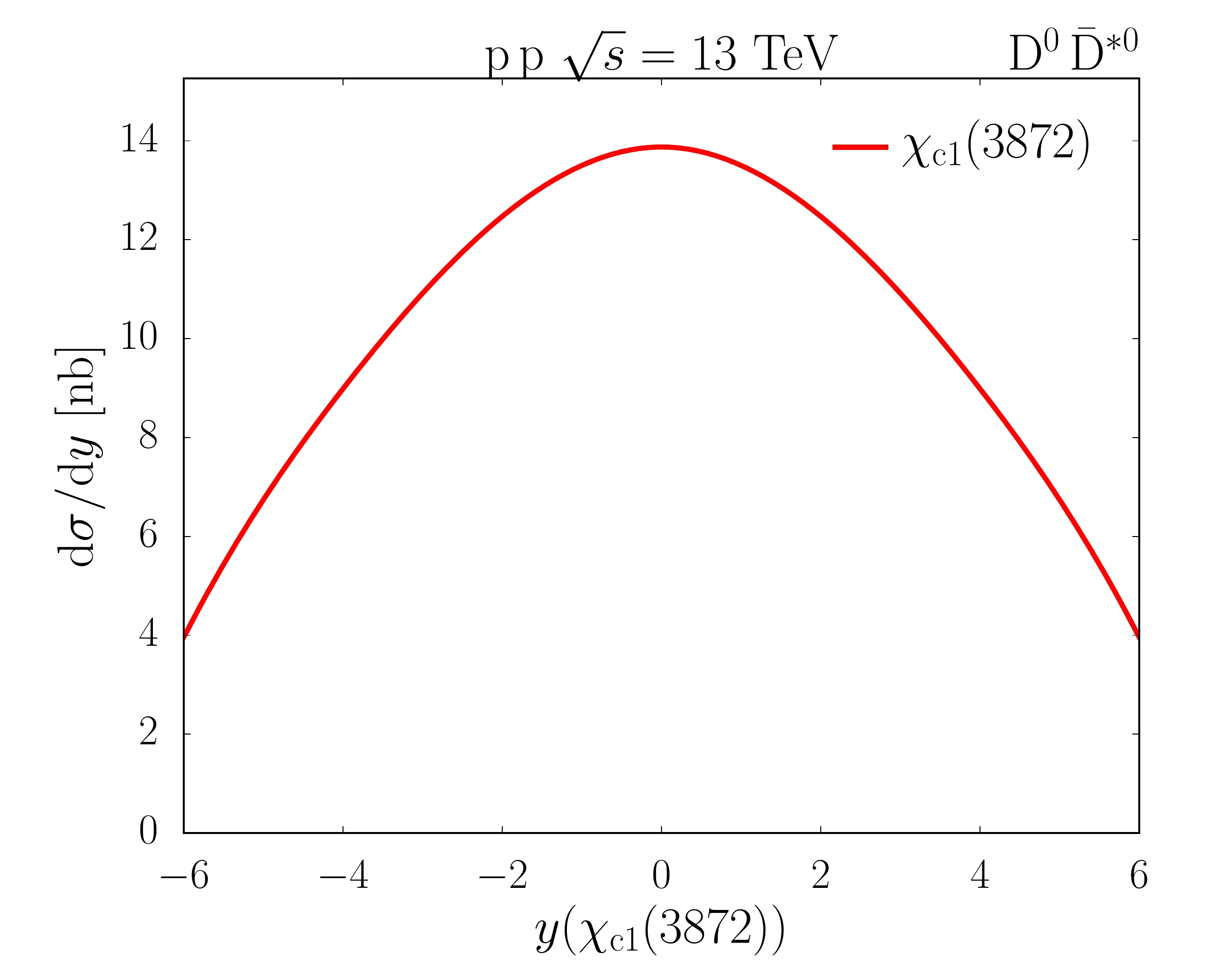} \plot{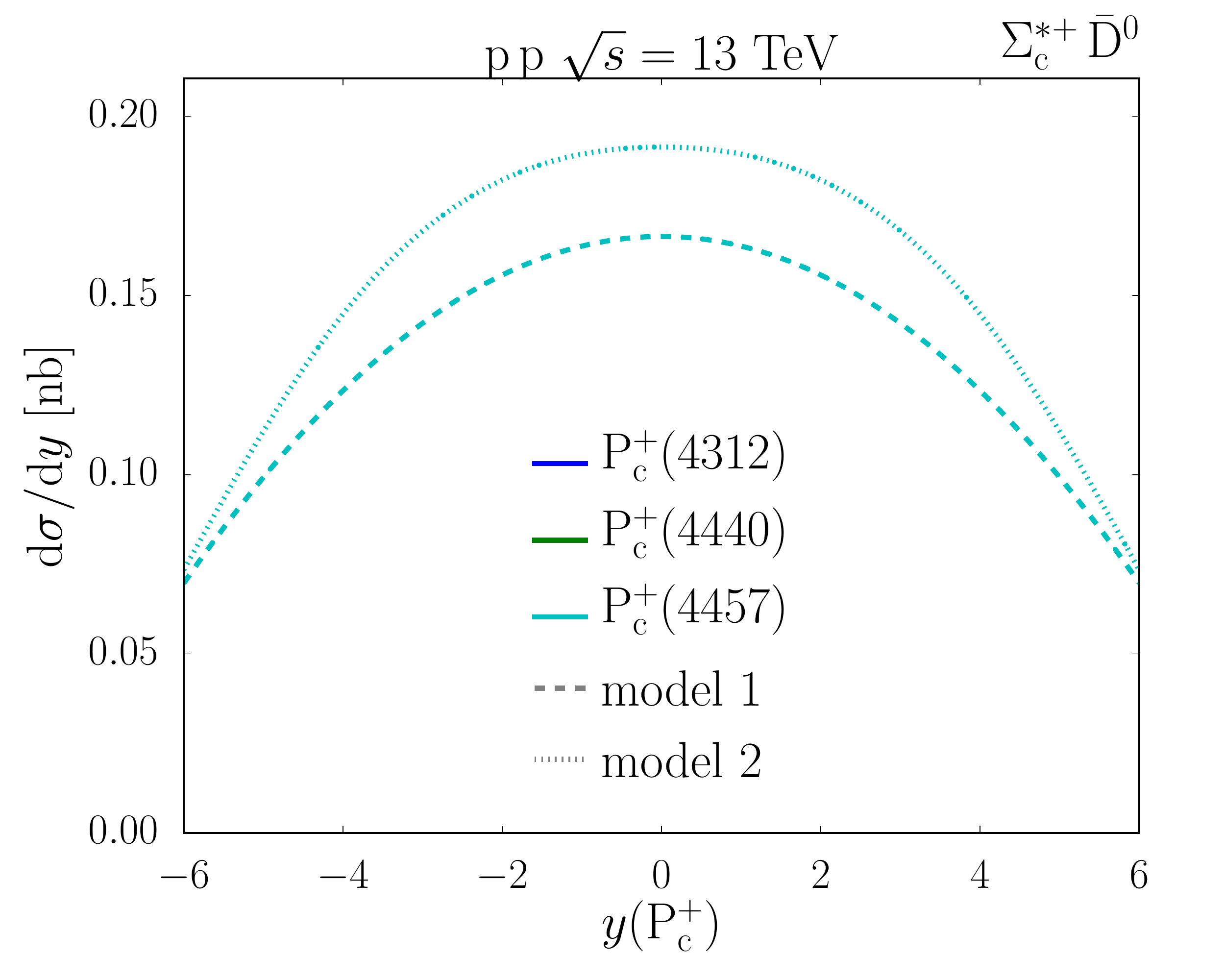}  \\
  \plot{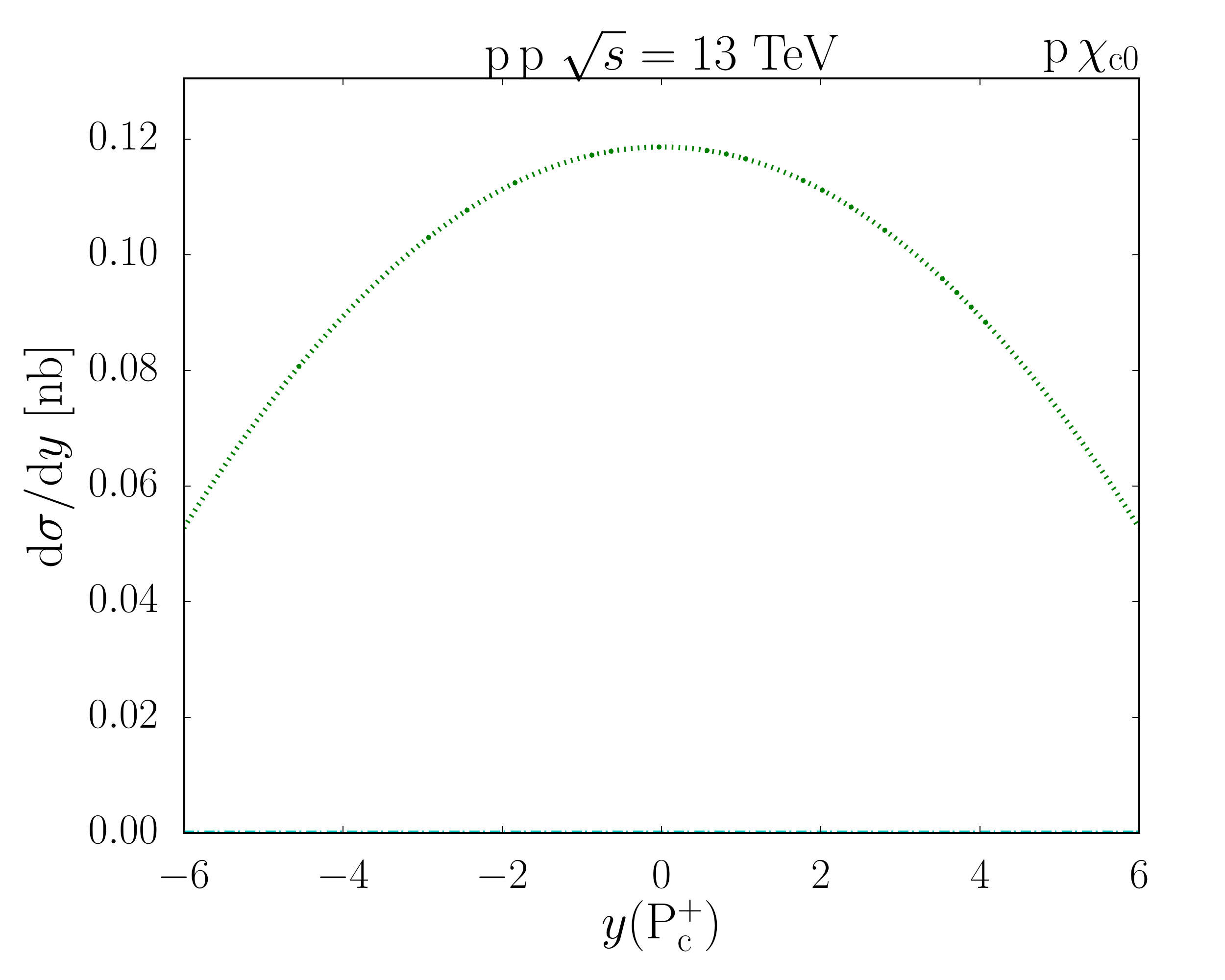}  \plot{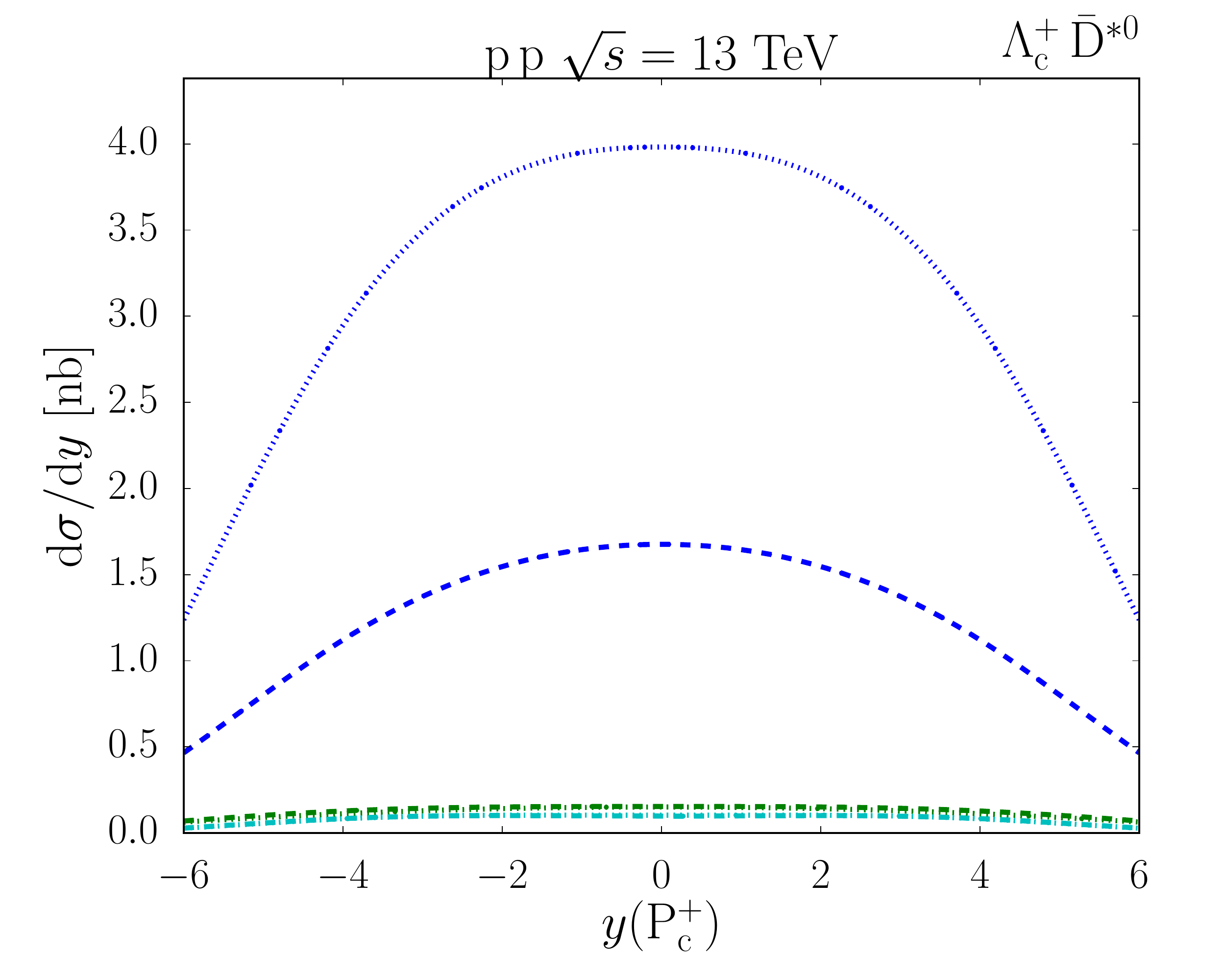} \\
  \caption{Tetraquark and pentaquark resonance hadronic-rescattering
    cross sections, differential in resonance $y$, produced in \pp
    collisions at $\sqrt{s} = 13~\TeV$.\label{fig:xsY}}
\end{figure}

The cross sections for tetraquark and pentaquark production at the LHC
with $\sqrt{s} = 13~\TeV$ are provided differentially in \pT and
rapidity by \figref{fig:xsPt} and \figref{fig:xsY}, respectively. The
cross sections are separated by rescattering channel, where the first
four channels of \figref{fig:xsECM} are shown. For the \Tc, only the
dominant \DDstar channel is given although both \Jpsiomega and
\Jpsirho channels do contribute, but at \bigo{10^{-2}} the rate of
\DDstar production. For the $\Pc(4312)$ and $\Pc(4440)$ resonances the
\LambdacDstar channel is the leading production mechanism, while the
\SigmacstarD channel is the dominant channel for $\Pc(4457)$
production. The \SigmacD channel contributes to $\Pc(4440)$ and
$\Pc(4457)$ production for both models, but is always subleading. The
channels with light baryons do not significantly contribute to
pentaquark production except the \pchic channel for the model 2
$\Pc(4440)$ state. Note that the \pJpsi discovery channel is not
relevant for \Pc production.

The \pT distributions of both the \Tc and \Pc states peak near
$2~\GeV$. However, the \pchic channel is significantly softer than the
other channels, peaking near $1~\GeV$. It is important to note that
the production of the \chic[0] in \Pythia is via a hard NRQCD matrix
element, \eg $\g\,\g \to \g\,\chic[0]$, where the low \pT divergence
has been regulated with a \pT damping term. This is in contrast to the
other hadrons, which are produced directly from the hadronization
process. The rapidity distribution for the \Tc peaks centrally as do
the \Pc rapidity distributions, although the \Pc distributions are
slightly broader.

\begin{figure}
  \centering
  \centering
  \plot{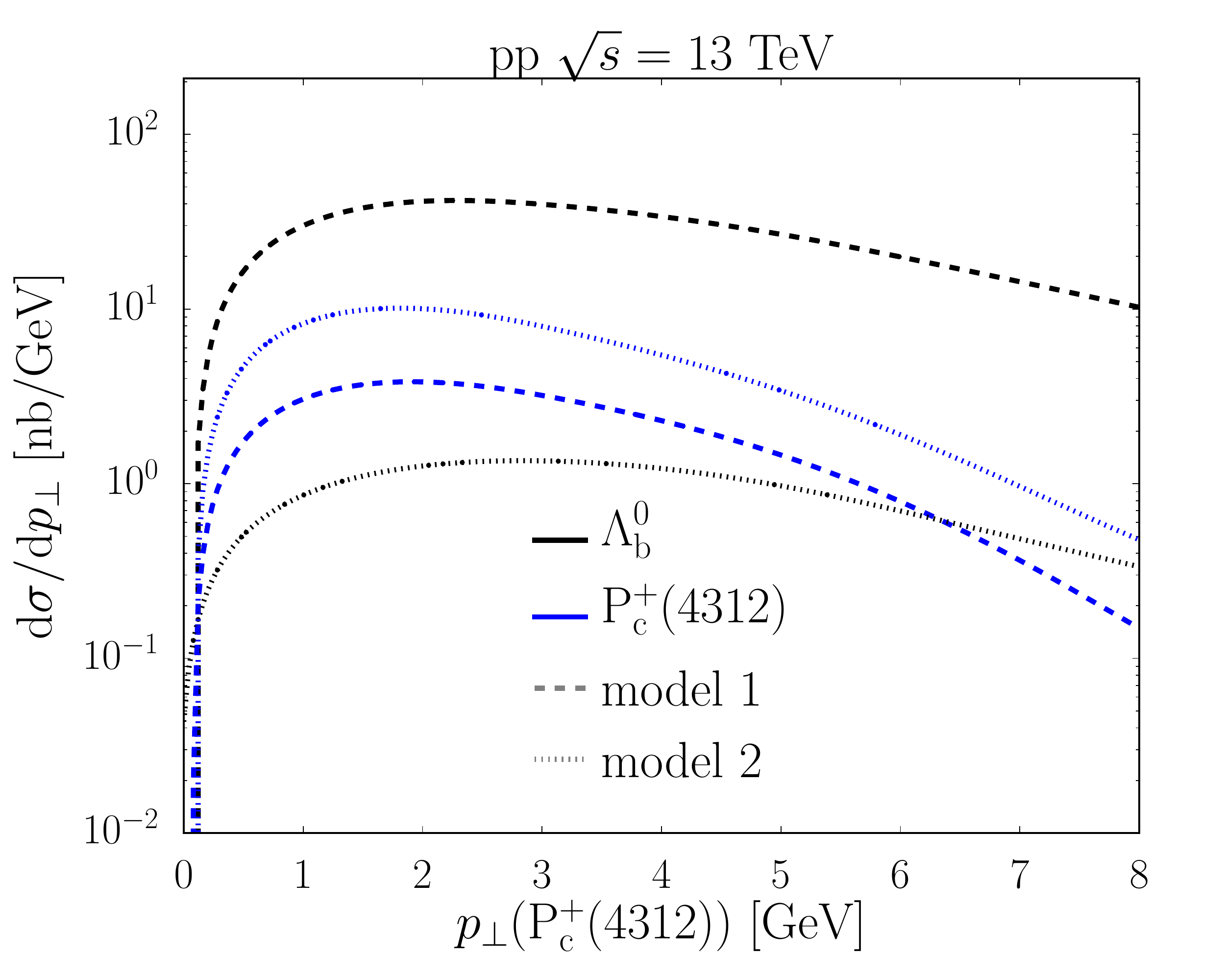} \plot{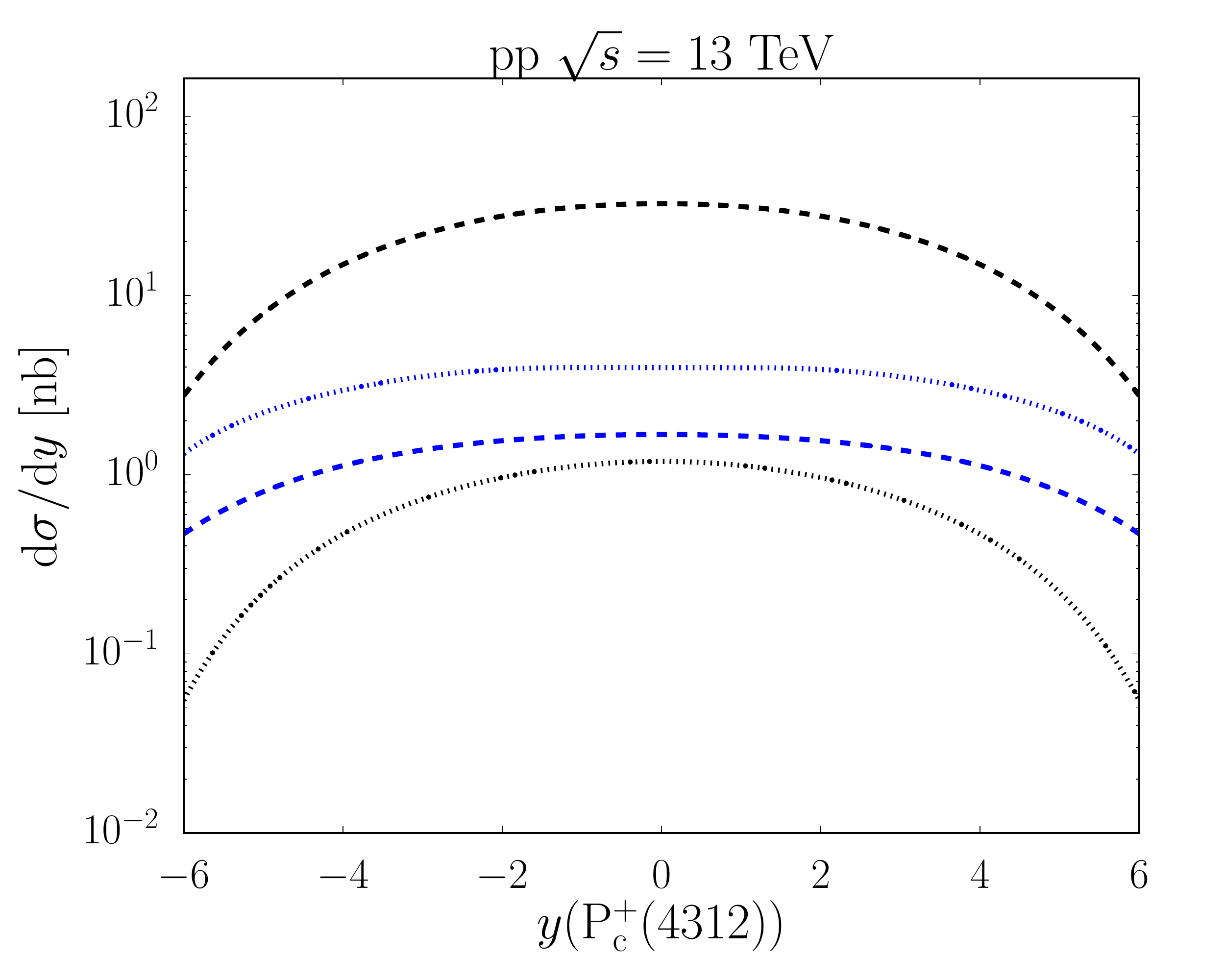} \\
  \plot{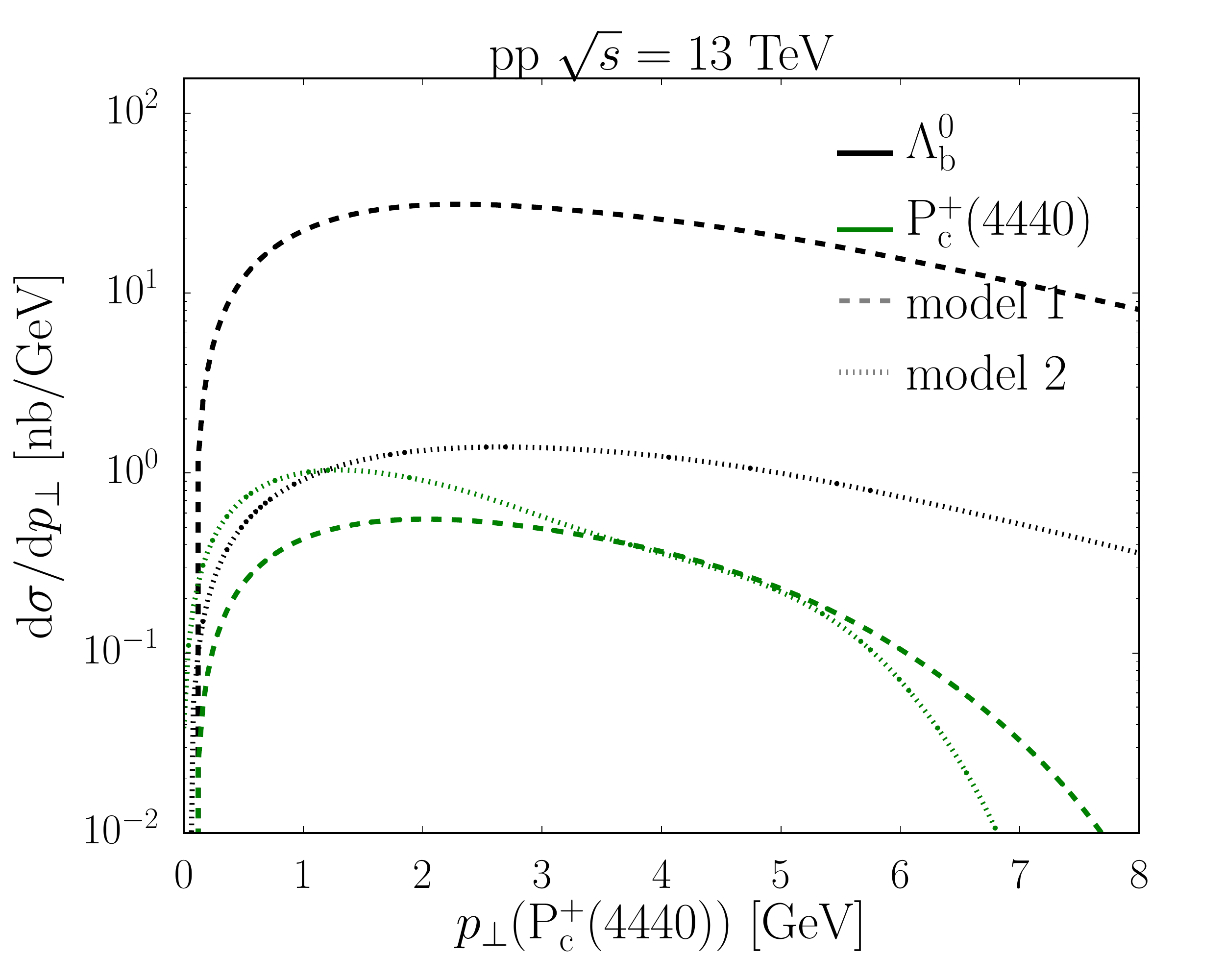} \plot{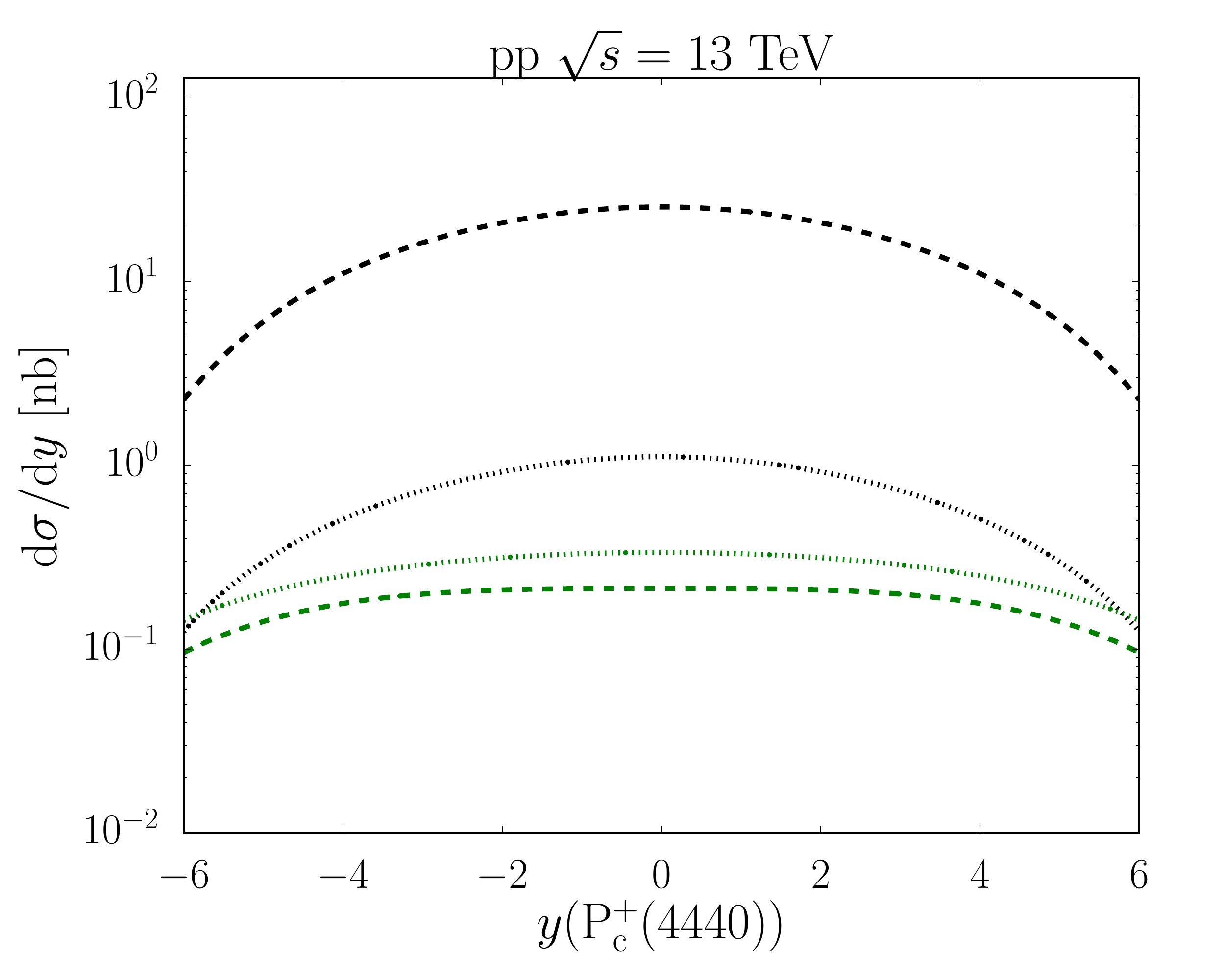} \\
  \plot{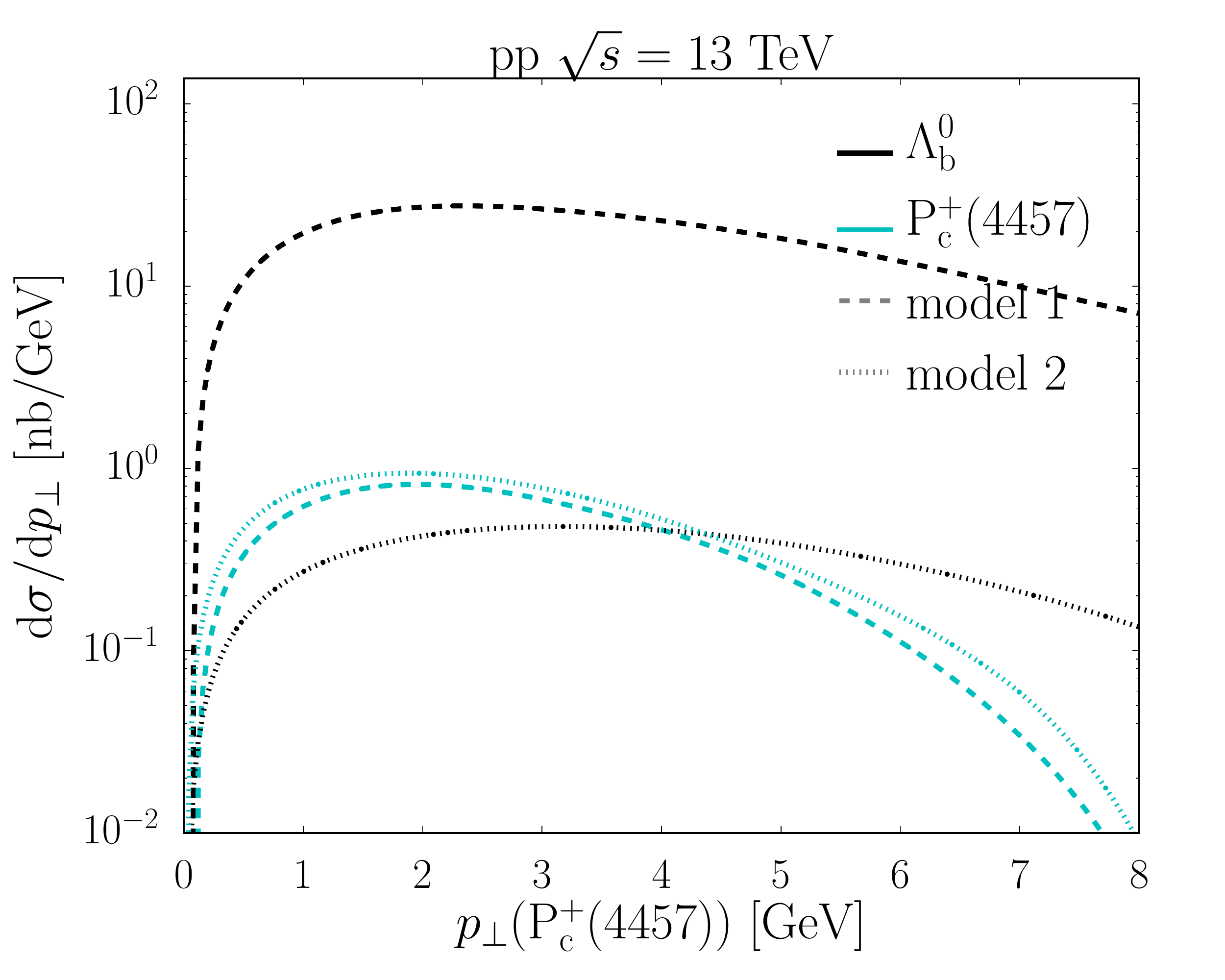} \plot{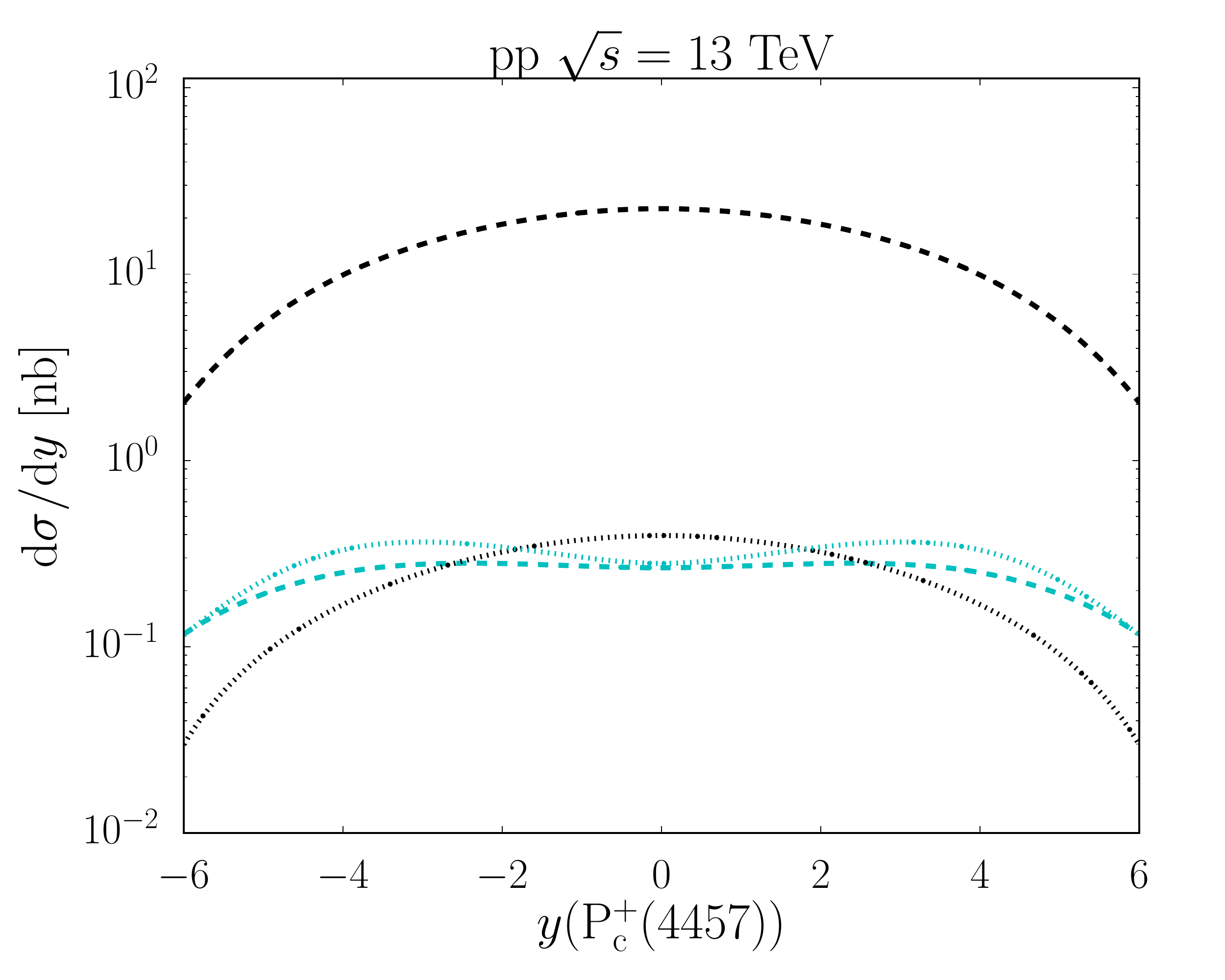} \\
  \caption{Total hadronic-rescattering cross-section for (dashed)
    model 1 and (dotted) model 2 pentaquarks, compared to pentaquark
    production from (black) \Lambdab decays.\label{fig:xsDec}}
\end{figure}

The total hadronic-rescattering cross sections for the pentaquarks are
compared to the expected cross section from \Lambdab decays in
\figref{fig:xsDec}. In general, the \pT distributions between hadronic
rescattering and \Lambdab decays are similar, and again tend to peak
around $2~\GeV$. For the model 2 $\Pc(4440)$ state, the \pchic
channel contributes at the same level as the \LambdacDstar
channel. This results in the softer $\Pc(4440)$ \pT spectrum of model
2 in comparison to model 1. The rapidity distributions for \Pc states
produced from \Lambdab decays are more central than for those produced
in rescattering. For the former, this primarily depends upon the
rapidity of the parent \Lambdab baryons, which in turn depend upon the
jets from which they are produced. For the latter, the distributions
are broader, and there is a rapidity dip around zero for the
$\Pc(4457)$ state.

This dip is particularly pronounced in channels with lighter hadrons,
\eg \ppi, whose total contribution to pentaquark formation is very
small overall. Similar dips also appear in the rapidity spectrum of
\ppi rescatterings with invariant masses around the pentaquark masses,
regardless of whether pentaquarks are actually formed. Hence, this is
a general feature of the rescattering framework, and not specific to
pentaquarks. Light hadrons must have a more pronounced difference in
momenta in order to reach the invariant masses necessary for
pentaquark formation. Consequently, the rapidity difference between
the rescattering particles should be non-zero, although this still
does not fully explain why the total momentum also has non-zero
rapidity. Since this is a matter of rescattering in general, and has
only a small effect on exotic hadrons, a more in-depth study is left
for the future.

The cross sections for \Pc from rescattering are generally similar
between the two models, which is expected since the dominant partial
widths remain similar between the models. The cross section from
\Lambdab decays differs significantly between the two models,
however. This is because the \Jpsi partial widths, used in determining
$\BR(\Lambdab \to \PcK)$, differ by \bigo{10} to \bigo{100}. This has
important experimental implications. Measuring the \Lambdab cross
section in conjunction with the $\Lambdab \to \PcK$ production can
help separate molecular state models, with expected differences as
large as \bigo{100}.

\subsection{Tetraquarks with LHCb}

The fiducial cross section for \Tc production in \pp collisions at a
centre-of-mass energy of $\sqrt{s} = 7~\TeV$ has been measured by
LHCb~\cite{LHCb:2011zzp} to be,
\begin{equation*}
  \sigma_\textrm{LHCb}(\pp \to \Tc[\to \Jpsi \, \di{\pi}] + X) =
  5.4 \pm 1.3 \pm 0.8~\nb
\end{equation*}
where the $\Tc \to \Jpsirho$ final state with a $\rho^0 \to \di{\pi}$
decay has been used. For this fiducial cross section, the
pseudorapidity of the \Tc is required to be in the range $2.0 \leq
\eta \leq 4.5$, and its momentum must be in the range $5
\leq \pT \leq 20~\GeV$. These requirements ensure efficient detection
of the final state and help minimise systematic uncertainties due to
particle reconstruction inefficiencies. This cross section does not
separate the prompt \Tc production from feed-down production, where
the \Tc is produced from the decay of a heavier hadron. Indeed, \Tc
production from \B-hadron decays is expected to be sizeable.

The predicted cross section for \Tc production from hadronic
rescattering in \pp collisions at $\sqrt{s} = 7~\TeV$ is,
\begin{equation*}
  \sigma_{\text{rescatter}}(\pp \to \Tc[\to \Jpsi \, \di{\pi}] + X) =
  0.04~\nb
\end{equation*}
where the uncertainty on this cross section will have contributions
from the uncertainty of the $\D^0/\Dbar^0$ and $\D^{*0}/\Dbar^{*0}$
meson fluxes, and the uncertainty of the estimated hadronic
rescattering cross sections. The former uncertainty depends upon the
tuning of \Pythia used, while the latter depends not only on the
modelling of the \Tc line-shape but also the partial widths for each
rescattering channel. Both of these uncertainties are difficult to
quantify and so the hadronic rescattering cross section is quoted here
without uncertainty, with the explicit understanding that the
uncertainty may be large.

The predicted hadronic-rescattering cross section is not larger than
the measured total cross section, which lends some credence to this
hadronic-rescattering model. However, the hadronic-rescattering cross
section is \bigo{100} times less than the measured cross section,
indicating that hadronic rescattering is expected to provide a
negligible contribution to \Tc production. Utilising lifetime
measurement capabilities, future LHCb measurements could separate the
\Tc cross-section into prompt and feed-down production, allowing for a
direct comparison with this prediction. This could help determine how
best to model prompt \Tc formation, whether from direct NRQCD
calculations, parton showers, hadronization, hadronic rescattering,
\etc

\subsection{Pentaquarks with LHCb}

While pentaquark production has been unambiguously observed by LHCb
using the exclusive $\Lambdab \to \pJpsi\,\K^-$ decay, no \Pc cross
sections from \Lambdab decays or otherwise, have been measured. To
fully understand the nature of the observed pentaquark states, these
measurements are necessary, including separate cross-section
measurements of prompt and feed-down pentaquark production. Depending
upon the expected rate of prompt pentaquark production, this may be
challenging, as there can be large combinatorial backgrounds when
considering a prompt pentaquark signal. The displaced vertex, $\K^-$
in the final state, and \Lambdab mass constraint are all no longer
available when searching for prompt pentaquark
candidates. Consequently, a clean and fully reconstructed final state
of charged particles is preferred to reduce these backgrounds.

For the model 1 $\Pc(4312)$, the largest branching fraction for an
exclusive charged final state is \bigo{10^{-4}} with at least five
charged particles, including two or more $\pi^\pm$. The combinatorial
backgrounds for such a final state will be large, and the high
multiplicity will also result in relatively low momentum charged
particles that are difficult to reconstruct. The $\petac [\to
  \di{\K}]$ final state is expected to have a branching ratio of
\bigo{10^{-5}}, while the $\pJpsi [\to \di{\mu}]$ branching ratio will
be \bigo{10^{-6}}. For the model 1 $\Pc(4440)$, the branching ratio
for the $\pJpsi [\to \di{\mu}]$ final state is \bigo{10^{-4}}, where
all fully charged final states with higher branching ratios up to
\bigo{10^{-3}} have multiplicities of five or larger. The situation
for the $\Pc(4457)$ is similar, but with the relevant branching ratios
reduced by an order of magnitude. Consequently, without detailed
background and detector simulation, the $\Pc \to \pJpsi [\to
  \di{\mu}]$ decay still provides a reasonable final state for the
model 1 pentaquarks.

The discovery channel of \pJpsi is enhanced in model 2 by \bigo{10} to
\bigo{100} with respect to model 1. For all model 2 pentaquarks, the
$\pJpsi [\to \di{\mu}]$ final state branching ratios are the same
order of magnitude as the leading branching ratios, which for the
$\Pc(4440)$ and $\Pc(4457)$ are in the $\prho [\to \di{\pi}]$
channel. The $\pJpsi [\to \di{\mu}]$ branching ratio is \bigo{10^{-4}}
for the $\Pc(4312)$ and \bigo{10^{-3}} for the $\Pc(4440)$ and
$\Pc(4457)$ states. The $\pJpsi [\to \di{\e}]$ branching ratios are
also the same as the $\pJpsi [\to \di{\mu}]$ branching ratios, but in
the context of LHCb, electron reconstruction and identification is
significantly more challenging than for muons.

This study considers using the LHCb detector to measure prompt
pentaquark production via hadronic rescattering. The \run{3} LHCb
detector~\cite{LHCb:2012doh, Bediaga:2013yyz, LHCb:2013lvx,
  LHCb:2014gmu, LHCbCollaboration:2014vzo, LHCb:2018roe} is a forward arm
spectrometer with full particle reconstruction between
pseudorapidities of $2$ and $5$, including a precision vertex
detector, a charged particle tracking system, Cherenkov detectors
providing particle identification, an electromagnetic calorimeter, and
a muon system. Additionally, during \run{3} the LHCb data acquisition
system will employ a real-time analysis strategy~\cite{LHCbCollaboration:2014vzo},
where the entire detector will be read out and calibrated in real
time. This will enable the full reconstruction of pentaquark
candidates during online data taking, and minimise possible
inefficiencies of data acquisition. The target \run{3} integrated
luminosity for LHCb of $15~\ifb$ at $\sqrt{s} = 13~\TeV$ is assumed
for this study~\cite{LHCb:2018roe}.

Selecting the $\pJpsi [\to \di{\mu}]$ final state, the following LHCb
fiducial requirements are made: the muons and protons must have $2 <
\eta < 5$; the muons must have $\pT > 0.5~\GeV$; and the proton must
have $\pT > 1~\GeV$. Given this fiducial selection, and assuming
similar performance to the \run{2} detector~\cite{LHCb:2014set}, the
reconstruction efficiency is expected to be near $100\%$. The
background is evaluated as combinations of prompt real $\Jpsi [\to
  \di{\mu}]$ decays with prompt protons. Background from particle
mis-identification will also contribute, but is expected to be
subleading. Real displaced \Pc signals can be separated from the
prompt \Pc signal with fits to lifetime observables.

\begin{figure}
  \plot{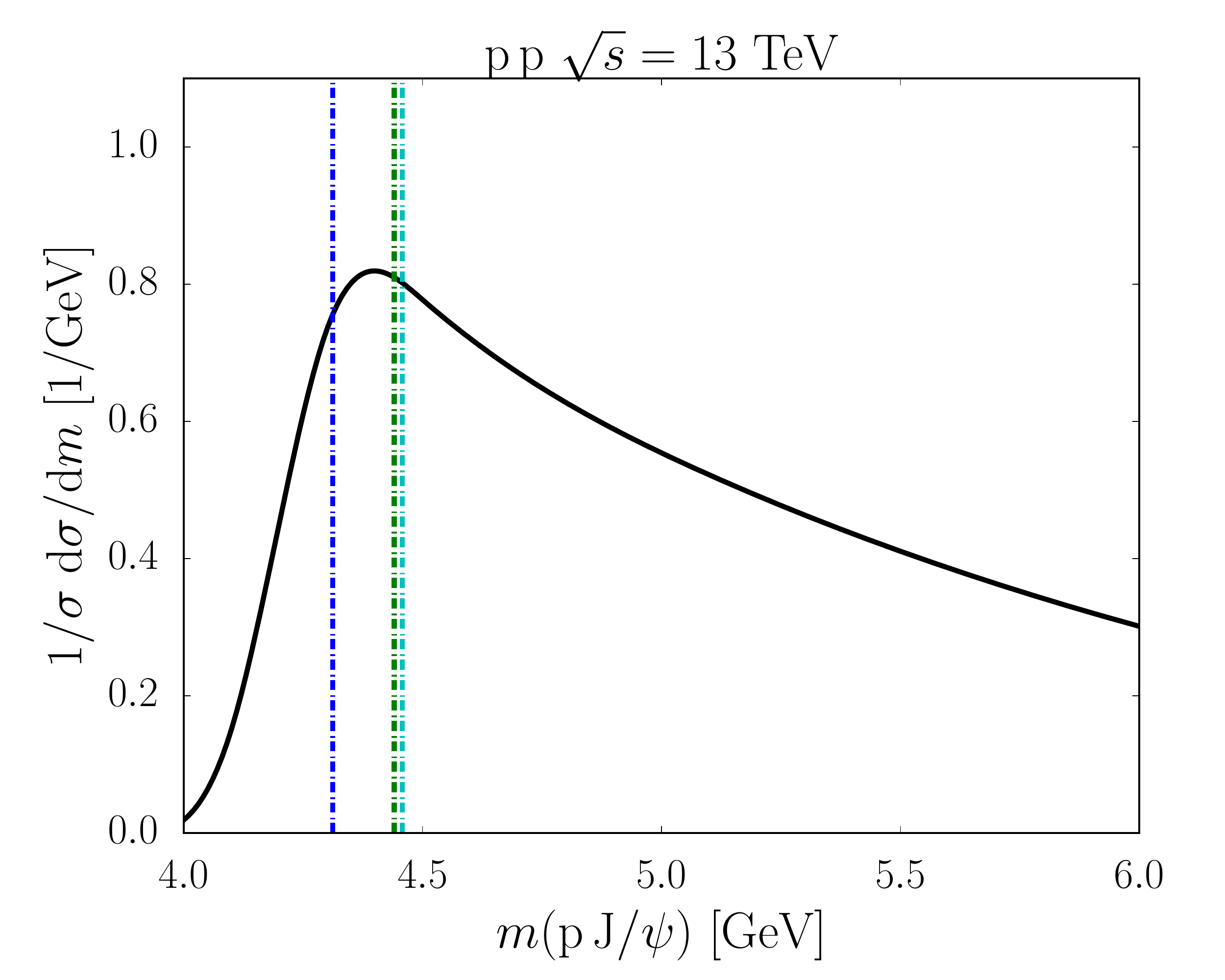}
  \plot{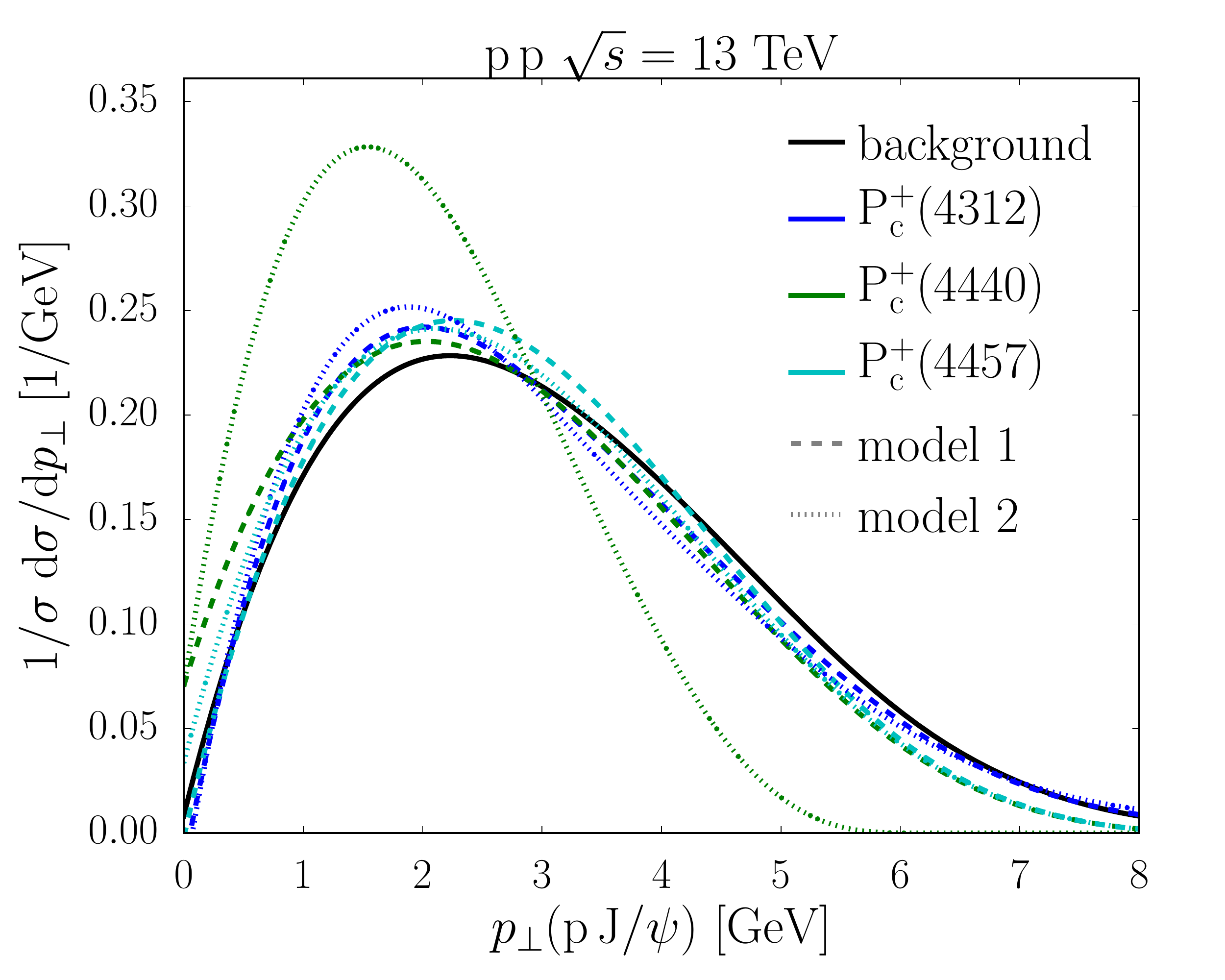} \\
  \plot{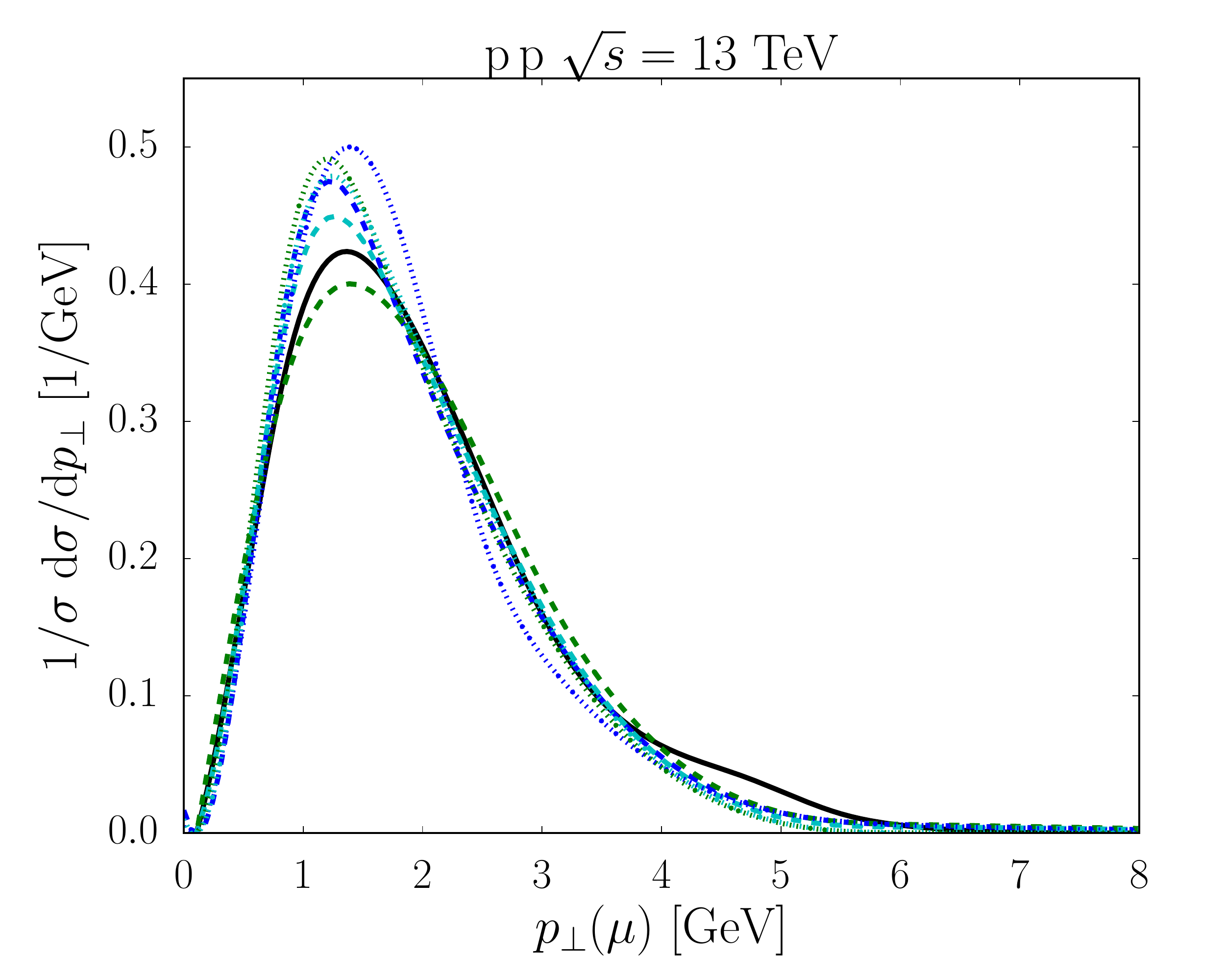}
  \plot{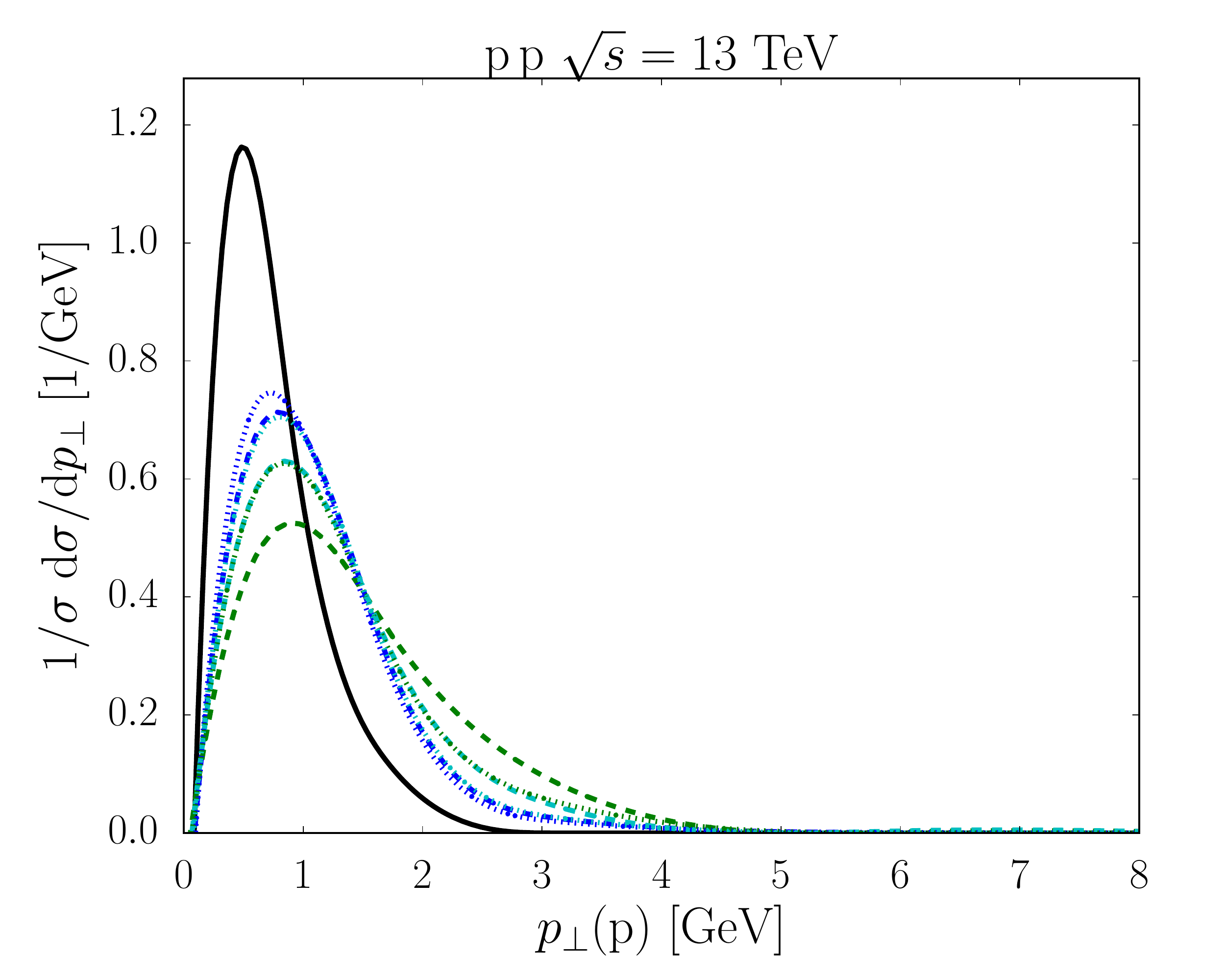} \\
  \caption{Normalised cross-sections for $\Pc \to \pJpsi$ decays and
    (solid black) combinatorial background, differential in (top left)
    final-state mass, (top right) final-state \pT, (bottom left)
    muon \pT, and (bottom right) proton \pT.\label{fig:bkg}}
\end{figure}

Comparisons of the background distributions with the signal
distributions are shown in \figref{fig:bkg} with only the final state
pseudorapidity requirements in place. All \pT distributions are
similar between the \Pc states, except for the model 2 $\Pc(4440)$
state, due to the contribution of the \pchic channel. The muon \pT is
slightly softer for the signal than for the background, while the
proton \pT is slightly harder for the signal than the
background. However, these differences are not sufficiently pronounced
to provide any kinematic discrimination. Further separation of signal
and background may be possible, but would require detailed detector
simulation more suitable within an experimental context. The \Pc
candidate mass distribution peaks near the nominal \Pc masses, which
complicates resonance fitting. Increasing the $\pT$ requirement on the
proton could flatten this distribution, but will reduce the signal
selection efficiency.

\begin{table}
  \centering
  \caption{Expected number of reconstructed prompt background and
    signal \Pc candidates by the LHCb detector during
    \run{3}.\label{tab:nPc}}
  \begin{tabular}{>{$}l<{$}|
    >{$}l<{$} >{$}l<{$} >{$}l<{$} >{$}l<{$}}
    \toprule
    & \mathrm{bkg}
    & \mathrm{P_c^+}(4312)
    & \mathrm{P_c^+}(4440)
    & \mathrm{P_c^+}(4457) \\
    \midrule
    \mbox{model 1}
    & \multirow{2}{*}{$\ef{2}{8}$}
    & \ef{2}{2}
    & \ef{2}{2}
    & \ef{1}{2} \\
    \mbox{model 2}
    &
    & \ef{1}{4}
    & \ef{5}{3}
    & \ef{8}{3} \\
    \bottomrule
  \end{tabular}
\end{table}

The LHCb \di{\mu} mass resolution is roughly $0.4\%$ of the dimuon
mass ~\cite{LHCb:2014set}. Even assuming twice the resolution for the
three-body \Pc candidate with the inclusion of the proton, this is
well below the natural width of the \Pc states. Consequently, a mass
window of roughly six times the \Pc total width is used when
determining the number of expected background candidates. The expected
number of signal and background candidates is given in
\tabref{tab:nPc}. The mass window of $4.4 < m < 4.5~\GeV$ is
considered for the background when comparing to all three \Pc
states. The number of background candidates will vary slightly given
different \Pc pole masses, but for the purposes of this study this
approximation is sufficient. From \tabref{tab:nPc} it is clear that
prompt \Pc production from hadronic rescattering, given the models
considered here, will not be observable by LHCb. However, even limits
on prompt \Pc production will still provide a valuable tool in
understanding the nature of the observed pentaquark states.

\section{Conclusion\label{sec:conclusion}}

The hadronic rescattering framework in \Pythia has been modified to
allow for the production of arbitrary hadronic resonances, with an
emphasis placed on the production of exotic hadrons that may be
molecular states. The relevant code will be published in an upcoming
\Pythia release. The production cross sections for the tetraquark
candidate \Tc, and the pentaquark candidates $\Pc(4312)$, $\Pc(4440)$,
and $\Pc(4457)$, have been calculated for $\sqrt{s} = 13~\TeV$ \pp
collisions at the LHC. Using this implementation of \Pythia, hadronic
rescattering predictions could also be made for the additional exotic
states discovered at the LHC. The hadronic rescattering cross section
for \Tc production at $\sqrt{s} = 7~\TeV$ in \pp collisions is
compared to the inclusive \Tc cross-section measurement by LHCb and
found to contribute at a $1\%$ level. Finally, the expected number of
prompt \Pc candidates from hadronic rescattering observed by LHCb
during \run{3}, using the exclusive final state of $\pJpsi[\to
  \di{\mu}]$, is estimated and found to be significantly smaller than
the estimated prompt background. However, cross-section measurements
of the \Pc candidates, separated into prompt and displaced
contributions, can still differentiate between predicted molecular
models of these pentaquarks.

\section*{Acknowledgements\label{sec:acknowledgements}}

The authors would like to thank Stephen Mrenna for initial discussions
about modelling pentaquark formation and Torbj\"orn Sj\"ostrand for
further general discussion. PI is supported by the U.S. National
Science Foundation grant OAC-2103889. MU is supported in part by the
Swedish Research Council, contract number 2016-05996, and in part by
the MCnetITN3 H2020 Marie Curie Innovative Training Network, grant
agreement 722104.

\bibliographystyle{utphys}
\bibliography{bibliography}

\end{document}